\documentclass[12pt,a4paper]{article}
\usepackage[dvips]{epsfig}
\begin{document}
%%%%%%%%%%%%%%%%%%%%%%%%%%%%%%%%%%%%%%%%%%%%%%%%%%%%%%%%%%%%%%%%%%%%%%%%
\date{}

\title{
{\vspace{-1em} \normalsize
\hfill \parbox{50mm}{DESY 99-029\\MS-TPI-99-02}}\\[25mm]
{\Large\bf Monte Carlo simulation of                         \\
SU(2) Yang-Mills theory with light gluinos}                  \\[8mm]}
\author{DESY-M\"unster Collaboration                         \\[8mm]
I.~Campos, R.~Kirchner, I.~Montvay, J.~Westphalen            \\
Deutsches Elektronen-Synchrotron DESY,                       \\
Notkestr.\,85, D-22603 Hamburg, Germany                      \\[5mm]
A.~Feo, S.~Luckmann, G.~M\"unster, K.~Spanderen              \\
Institut f\"ur Theoretische Physik I,                        \\
Universit\"at M\"unster, Wilhelm-Klemm-Str.\,9,               \\
D-48149 M\"unster, Germany}

%%%%%%%%%%%%%%%%%%%%%%%%%%%%%%%%%%%%%%%%%%%%%%%%%%%%%%%%%%%%%%%%%%%%%%%%
\newcommand{\be}{\begin{equation}}
\newcommand{\ee}{\end{equation}}
\newcommand{\half}{\frac{1}{2}}
\newcommand{\rar}{\rightarrow}
\newcommand{\lar}{\leftarrow}
%%%%%%%%%%%%%%%%%%%%%%%%%%%%%%%%%%%%%%%%%%%%%%%%%%%%%%%%%%%%%%%%%%%%%%%%

\maketitle
\vspace{3em}

\begin{abstract} \normalsize
 In a numerical Monte Carlo simulation of SU(2) Yang-Mills theory with
 light dynamical gluinos the low energy features of the dynamics as
 confinement and bound state mass spectrum are investigated.
 The motivation is supersymmetry at vanishing gluino mass.
 The performance of the applied two-step multi-bosonic dynamical fermion
 algorithm is discussed.
\end{abstract}

%%%%%%%%%%%%%%%%%%%%%%%%%%%%%%%%%%%%%%%%%%%%%%%%%%%%%%%%%%%%%%%%%%%%%%%%
\newpage
\section{Introduction}\label{sec1}

 Supersymmetry seems to be a necessary ingredient of a quantum theory
 of gravity.
 It is generally assumed that the scale where supersymmetry becomes
 manifest is near to the presently explored electroweak scale and that
 the supersymmetry breaking is spontaneous.
 An attractive possibility for spontaneous supersymmetry breaking is to
 exploit non-perturbative mechanisms in supersymmetric gauge theories.
 Therefore the non-perturbative study of supersymmetric gauge theories
 is highly interesting \cite{AKMRV}.
 In recent years there has been great progress in this field, in
 particular following the seminal papers of Seiberg and Witten
 \cite{SEIWIT}.

 The simplest supersymmetric gauge theories are the $N=1$ supersymmetric
 Yang-Mills (SYM) theories.
 Besides the gauge fields they contain massless Majorana fermions in the
 adjoint representation, which are called gauginos in general.
 In the context of strong interactions one can call the gauge fields
 {\em gluons} and the gauginos {\em gluinos}.
 In the simple case of a gauge group ${\rm SU}(N_c)$ the adjoint
 representation is $(N_c^2-1)$-dimensional, hence there are $(N_c^2-1)$
 gluons and the same number of gluinos.

 The basic assumption about the non-perturbative dynamics of SYM
 theories is that there is {\em confinement} and {\em spontaneous chiral
 symmetry breaking}, similar to QCD.
 The confinement is realized by colourless bound states.
 Their mass spectrum is supposed to show a non-vanishing lower bound -
 the {\em mass gap}.
 Since external colour sources in the fundamental representation cannot
 be screened, the asymptotics of the static potential is characterized
 by a non-vanishing {\em string tension}.
 The expected pattern of spontaneous chiral symmetry breaking in SYM
 theories is quite peculiar: considering for definiteness the gauge
 group SU($N_c$), the expected symmetry breaking is $Z_{2N_c} \to Z_2$.
 For this we have recently found a first numerical evidence in a Monte
 Carlo simulation \cite{PHASETRANS}.
 These general features of the low energy dynamics can be summarized
 in terms of low energy effective actions \cite{VENYAN,FAGASCH}.

 The supersymmetric point in the parameter space corresponds to
 vanishing gluino mass ($m_{\tilde{g}}=0$).
 For non-zero gluino mass the supersymmetry is {\em softly broken} and
 the physical quantities like masses, string tension etc. are supposed
 to be analytic functions of $m_{\tilde{g}}$.
 The linear terms of a Taylor expansion in $m_{\tilde{g}}$ are often
 determined by the symmetries of the low energy effective actions
 \cite{LINEARMASS}.

 The lattice regularization offers a unique possibility to confront
 the expected low energy dynamical features of supersymmetric gauge
 theories with numerical simulation results (for a recent review see
 \cite{EDINBURGH}.)
 On the lattice it is natural to extend the investigations to a general
 value of the gluino mass.
 In fact, to study exactly zero gluino mass is usually more difficult
 than the massive case and often an extrapolation to the massless point
 is necessary.
 The main difficulty in the numerical simulations is the inclusion
 of light dynamical gluinos.
 Although one can gain some insight also by studies of the quenched
 theory \cite{QUENCHED,DGHV}, the supersymmetry requires dynamical
 light gluinos.

 In the present paper we report on a first large scale numerical
 investigation of SU(2) SYM theory with light gluinos.
 Although some preliminary results have already been published
 previously on different occasions
 \cite{STLOUIS,DESYMUNSTER,BOULDER,SPANDEREN} and the question of the
 discrete chiral symmetry breaking has been dealt with in a recent
 letter \cite{PHASETRANS}, this is the first detailed presentation of
 the obtained results.
 The numerical Monte Carlo simulations presented here have been
 performed on the CRAY-T3E computers at John von Neumann Institute for
 Computing (NIC), J\"ulich.

%%%%%%%%%%%%%%%%%%%%%%%%%%%%%%%%%%%%%%%%%%%%%%%%%%%%%%%%%%%%%%%%%%%%%%%%
\subsection{Lattice formulation}\label{sec1.1}

 For the lattice formulation we take the Wilson action, both for the
 gluon and gluino, as suggested some time ago by Curci and Veneziano
 \cite{CURVEN}.
 The effective gauge field action is
\be\label{eq1.1_1}
S_{CV} = \beta\sum_{pl} \left( 1-\half{\rm Tr\,}U_{pl} \right)
- \half\log\det Q[U] \ .
\ee
 For the gauge group ${\rm SU}(N_c)$, the bare gauge coupling is given
 by $\beta \equiv 2N_c/g^2$.
 The {\em fermion matrix} for the gluino $Q$ is defined by
\be  \label{eq1.1_2}
Q_{yv,xu} \equiv Q_{yv,xu}[U] \equiv
\delta_{yx}\delta_{vu} - K \sum_{\mu=1}^4 \left[
\delta_{y,x+\hat{\mu}}(1+\gamma_\mu) V_{vu,x\mu} +
\delta_{y+\hat{\mu},x}(1-\gamma_\mu) V^T_{vu,y\mu} \right] \ .
\ee
 $K$ is the hopping parameter and the matrix for the gauge-field link
 in the adjoint representation is defined as
\be  \label{eq1.1_3}
V_{rs,x\mu} \equiv V_{rs,x\mu}[U] \equiv
2 {\rm Tr}(U_{x\mu}^\dagger T_r U_{x\mu} T_s)
= V_{rs,x\mu}^* =V_{rs,x\mu}^{-1T} \ .
\ee
 The generators $T_r \equiv \half \lambda_r$ satisfy the usual
 normalization ${\rm Tr\,}(\lambda_r\lambda_s)=\half\delta_{rs}$.
 In case of SU(2) we have $T_r \equiv \half \tau_r$ with the isospin
 Pauli-matrices $\tau_r$.

 The fermion matrix for the gluino $Q$ in (\ref{eq1.1_2}) is not
 hermitean but it satisfies
\be  \label{eq1.1_4}
Q^\dagger = \gamma_5 Q \gamma_5 \ .
\ee
 This relation allows for the definition of the {\em hermitean fermion
 matrix}
\be  \label{eq1.1_5}
\tilde{Q} \equiv \gamma_5 Q = \tilde{Q}^\dagger \ .
\ee

 The factor $\half$ in front of $\log\det Q$ in (\ref{eq1.1_1}) tells
 that we effectively have a flavour number $N_f=\half$ of adjoint
 fermions.
 This describes Majorana fermions in the Euclidean path integral.
 For Majorana fermions the Grassmannian variables $\Psi_x$ and
 $\overline{\Psi}_x$ are not independent but satisfy, with the
 charge-conjugation Dirac matrix $C$,
\be\label{eq1.1_6}
\Psi_x = C \overline{\Psi}_x^T \ , \hspace{2em}
\overline{\Psi}_x = \Psi_x^T C \ .
\ee
 (Note that here we use the Dirac-Majorana field $\Psi_x$ instead of
 the Weyl-Majorana one $\lambda_x$.)
 The Grassmannian path integral for Majorana fermions is defined as
\be  \label{eq1.1_7}
\int [d\Psi] e^{ -\half\overline{\Psi} Q \Psi }
= \int [d\Psi] e^{ -\half\Psi^T CQ\Psi }
= {\rm Pf}(CQ) = {\rm Pf}(M) \ .
\ee
 Here the {\em Pfaffian} of the antisymmetric matrix $M \equiv CQ$
 is introduced.
 The Pfaffian can be defined for a general complex antisymmetric matrix
 $M_{\alpha\beta}=-M_{\beta\alpha}$ with an even number of dimensions
 ($1 \leq \alpha,\beta \leq 2N$) by a Grassmann integral as
\be  \label{eq1.1_8}
{\rm Pf}(M) \equiv
\int [d\phi] e^{-\half\phi_\alpha M_{\alpha\beta} \phi_\beta}
= \frac{1}{N!2^N} \epsilon_{\alpha_1\beta_1 \ldots \alpha_N\beta_N}
M_{\alpha_1\beta_1} \ldots M_{\alpha_N\beta_N} \ .
\ee
 Here, of course, $[d\phi] \equiv d\phi_{2N} \ldots d\phi_1$, and
 $\epsilon$ is the totally antisymmetric unit tensor.
 It can be easily shown that
\be  \label{eq1.1_9}
\left[{\rm Pf}(M)\right]^2 = \det M \ .
\ee

 Besides the partition function in (\ref{eq1.1_7}), expectation values
 for Majorana fermions can also be similarly defined
 \cite{TWO-STEP,DGHV}.
 It is easy to show that the hermitean fermion matrix for the gluino
 $\tilde{Q}$ has doubly degenerate real eigenvalues, therefore
 $\det Q = \det\tilde{Q} = \det M$ is positive and ${\rm Pf}(M)$ is
 real.
 In the effective gauge field action (\ref{eq1.1_1}) the absolute value
 of the Pfaffian is taken into account.
 The omitted sign can be included by reweighting the expectation values
 according to
\be\label{eq1.1_10}
\langle A \rangle = \frac{\langle A\; {\rm sign\,Pf}(M)\rangle_{CV}}
{\langle {\rm sign\,Pf}(M)\rangle_{CV}} \ .
\ee
 Here $\langle\ldots\rangle_{CV}$ means expectation value with respect
 to $S_{CV}$.
 This sign problem is very similar to the one in QCD with an odd
 number of quark flavours.

 The numerical simulations are almost always done on lattices with
 toroidal boundary conditions.
 In the three spatial directions it is preferable to take periodic
 boundary conditions both for the gauge field and the gluino.
 This implies that in the Hilbert space of states the supersymmetry is
 not broken by the boundary conditions.
 In the time direction in most cases we decided to choose periodic
 boundary conditions for bosons and antiperiodic ones for fermions,
 which is obtained if one writes traces in terms of Grassmann integrals:
 the minus sign for fermions is the usual one associated with closed
 fermion loops.

%%%%%%%%%%%%%%%%%%%%%%%%%%%%%%%%%%%%%%%%%%%%%%%%%%%%%%%%%%%%%%%%%%%%%%%%
\subsection{Overview}\label{sec1.2}

 The aim of this paper is to give a complete presentation of the methods
 we used and to report on our numerical results.
 We concentrate on the confinement features as mass spectrum, emergence
 of supersymmetric multiplets of bound states and string tension.

 The plan of the paper is as follows: in the next section the numerical
 simulation algorithm is described.
 In particular, the computation of the necessary Pfaffians is dealt with
 in section \ref{sec2.3}.
 The choice of algorithmic parameters and the observed autocorrelations
 are collected in section \ref{sec2.5}.
 The numerical results for the confinement potential and string tension
 as a function of the bare gluino mass are summarized in section
 \ref{sec3}.
 Section \ref{sec4} is devoted to the spectrum of bound states:
 glueballs, gluino-glueballs and gluinoballs.
 This also includes the questions about possible mixing (section
 \ref{sec4.4}).
 The last section contains a summary.
 Three appendices are included: appendix~\ref{appenA} about least squares
 optimized polynomials, appendix~\ref{appenB} on the methods used for
 the determination of the smallest eigenvalues and eigenvectors of the
 gluino matrix and appendix~\ref{appenC} about the main features of
 the C++ implementation of our computer code.

%%%%%%%%%%%%%%%%%%%%%%%%%%%%%%%%%%%%%%%%%%%%%%%%%%%%%%%%%%%%%%%%%%%%%%%%
\section{Multi-bosonic algorithm with corrections}\label{sec2}

 The multi-bosonic algorithm for Monte Carlo simulations of fermions has
 been proposed by L\"uscher \cite{LUSCHER}.
 In the original version for $N_f$ (Dirac-) fermion flavours one
 considers the approximation of the fermion determinant
\be\label{eq2_1}
\left|\det(Q)\right|^{N_f} =
\left\{\det(Q^\dagger Q) \right\}^{N_f/2}
\simeq \frac{1}{\det P_n(Q^\dagger Q)} \ ,
\ee
 where the polynomial $P_n$ satisfies
\be\label{eq2_2}
\lim_{n \to \infty} P_n(x) = x^{-N_f/2}
\ee
 in an interval $[\epsilon,\lambda]$ covering the spectrum of
 $Q^\dagger Q$.
 Note that here the absolute value of the determinant is taken which
 leaves out its sign (or phase).
 In case of $N_f=\half$, which corresponds to a Majorana fermion, this
 sign problem will be considered in section \ref{sec2.3}.

 For the multi-bosonic representation of the determinant one uses
 the roots of the polynomial $r_j,\; (j=1,\ldots,n)$
\be\label{eq2_3}
P_n(Q^\dagger Q) = P_n(\tilde{Q}^2) =
r_0 \prod_{j=1}^n (\tilde{Q}^2 - r_j) \ .
\ee
 Assuming that the roots occur in complex conjugate pairs, one can
 introduce the equivalent forms
\be\label{eq2_4}
P_n(\tilde{Q}^2)
= r_0 \prod_{j=1}^n [(\tilde{Q} \pm \mu_j)^2 + \nu_j^2]
= r_0 \prod_{j=1}^n (\tilde{Q}-\rho_j^*) (\tilde{Q}-\rho_j)
\ee
 where $r_j \equiv (\mu_j+i\nu_j)^2$ and $\rho_j \equiv \mu_j + i\nu_j$.
 With the help of complex scalar (pseudofermion) fields $\Phi_{jx}$
 one can write
\be\label{eq2_5}
\prod_{j=1}^n\det[(\tilde{Q}-\rho_j^*) (\tilde{Q}-\rho_j)]^{-1} \propto
\int [d\Phi]\; \exp\left\{ -\sum_{j=1}^n \sum_{xy}
\Phi_{jy}^+\, [(\tilde{Q}-\rho_j^*) (\tilde{Q}-\rho_j)]_{yx}\,
\Phi_{jx} \right\} \ .
\ee

 Since for a finite polynomial of order $n$ the approximation in
 (\ref{eq2_2}) is not exact, in principle, one has to extrapolate the
 results to $n\to\infty$.
 In practice this can also be done by investigating the $n$-dependence
 and showing that the systematic errors introduced by the finiteness of
 $n$ are negligible compared to the statistical errors.

 The difficulty for small fermion masses in large physical volumes is
 that the {\em condition number} $\lambda/\epsilon$ becomes very large
 ($10^4-10^6$) and very high orders $n = {\cal O}(10^3)$ are needed for
 a good approximation.
 This requires large storage and the autocorrelation becomes bad since
 it is proportional to $n$.
 One can achieve substantial improvements on both these problems by
 introducing a two-step polynomial approximation
 \cite{TWO-STEP,POLYNOM}.
 In this {\em two-step multi-bosonic} scheme (\ref{eq2_2}) is replaced
 by
\be\label{eq2_6}
\lim_{n_2 \to \infty} P^{(1)}_{n_1}(x)P^{(2)}_{n_2}(x) =
x^{-N_f/2} \ , \hspace{1em}
x \in [\epsilon,\lambda] \ .
\ee
 The multi-bosonic representation is only used for the first
 polynomial $P^{(1)}_{n_1}$ which provides a first crude approximation
 and hence the order $n_1$ can remain relatively low.
 The correction factor $P^{(2)}_{n_2}$ is realized in a stochastic
 {\em noisy correction step} with a global accept-reject condition
 during the updating process (see section \ref{sec2.1}).
 In order to obtain an exact algorithm one has to consider in this case
 the limit $n_2\to\infty$.
 For very small fermion (i.e.\ gluino) masses it turned out more
 practicable to fix some large $n_2$ and perform another small
 correction in the evaluation of expectation values by {\em reweighting}
 with a still finer polynomial (see section \ref{sec2.2}).

%%%%%%%%%%%%%%%%%%%%%%%%%%%%%%%%%%%%%%%%%%%%%%%%%%%%%%%%%%%%%%%%%%%%%%%%
\subsection{Update correction: global accept-reject}\label{sec2.1}

 The idea to use a stochastic correction step in the updating
 \cite{KENKUT}, instead of taking very large polynomial orders $n$, was
 proposed in the case of $N_f=2$ flavours in \cite{BOFOGA}.
 $N_f=2$ is special because the function to be approximated is just
 $x^{-1}$ and $P^{(2)}_{n_2}(x)$ can be replaced by the calculation of
 the inverse of $xP^{(1)}_{n_1}(x)$.
 For general $N_f$ one can take the two-step approximation scheme
 introduced in \cite{TWO-STEP}.

 The two-step multi-bosonic algorithm is described in detail in
 \cite{TWO-STEP}.
 Here we shortly repeat its main steps for the readers convenience
 and discuss the experience we obtained with it.
 The theory of the necessary optimized polynomials is summarized in
 appendix \ref{appenA} following \cite{POLYNOM}.

 In the two-step approximation scheme for $N_f$ flavours of fermions
 the absolute value of the determinant is represented as
\be\label{eq2.1_1}
\left|\det(Q)\right|^{N_f} \;\simeq\;
\frac{1}{\det P^{(1)}_{n_1}(\tilde{Q}^2)
\det P^{(2)}_{n_2}(\tilde{Q}^2)} \ .
\ee
 The multi-bosonic updating with $n_1$ scalar pseudofermion fields
 is performed by heatbath and overrelaxation sweeps for the scalar
 fields and Metropolis sweeps for the gauge field.
 After a Metropolis sweep for the gauge field a global accept-reject
 step is introduced in order to reach the distribution of gauge field
 variables $[U]$ corresponding to the right hand side of
 (\ref{eq2.1_1}).
 The idea of the noisy correction is to generate a random vector
 $\eta$ according to the normalized Gaussian distribution
\be \label{eq2.1_2}
\frac{e^{-\eta^\dagger P^{(2)}_{n_2}(\tilde{Q}[U]^2)\eta}}
{\int [d\eta] e^{-\eta^\dagger P^{(2)}_{n_2}(\tilde{Q}[U]^2)\eta}}  \ ,
\ee
 and to accept the change $[U^\prime] \lar [U]$ with probability
\be \label{eq2.1_3}
\min\left\{ 1,A(\eta;[U^\prime] \lar [U]) \right\} \ ,
\ee
 where
\be \label{eq2.1_4}
A(\eta;[U^\prime] \lar [U]) =
\exp\left\{-\eta^\dagger P^{(2)}_{n_2}(\tilde{Q}[U^\prime]^2)\eta -
            \eta^\dagger P^{(2)}_{n_2}(\tilde{Q}[U]^2)\eta\right\}\ .
\ee

 The Gaussian noise vector $\eta$ can be obtained from $\eta^\prime$
 distributed according to the simple Gaussian distribution
\be \label{eq2.1_5}
\frac{e^{-\eta^{\prime\dagger}\eta^\prime}}
{\int [d\eta^\prime] e^{-\eta^{\prime\dagger}\eta^\prime}}
\ee
 by setting it equal to
\be \label{eq2.1_6}
\eta = P^{(2)}_{n_2}(\tilde{Q}[U]^2)^{-\half} \eta^\prime  \ .
\ee

 In order to obtain the inverse square root on the right hand side of
 (\ref{eq2.1_6}), we can proceed with polynomial approximations in two
 different ways.
 The first possibility was proposed in \cite{TWO-STEP} with
 $x \equiv \tilde{Q}^2$ as
\be \label{eq2.1_7}
P^{(2)}_{n_2}(x)^{-\half} \simeq R_{n_3}(x) \simeq x^{N_f/4}
S_{n_s}[P^{(1)}_{n_1}(x)] \ .
\ee
 Here
\be \label{eq2.1_8}
S_{n_s}(P) \simeq  P^\half
\ee
 is an approximation of the function $P^\half$ on the interval
 $P \in [\lambda^{-N_f/2},\epsilon^{-N_f/2}]$.
 The polynomial approximations $R_{n_3}$ and $S_{n_s}$ can be determined
 by the same general procedure as $P^{(1)}_{n_1}$ and $P^{(2)}_{n_2}$.
 It turns out that these approximations are ``easier'' in the sense that
 for a given order higher precisions can be achieved than, say, for
 $P^{(1)}_{n_1}$.

 Another possibility to obtain a suitable approximation for
 (\ref{eq2.1_6}) is to use the second decomposition in (\ref{eq2_4})
 and define
\be\label{eq2.1_9}
P^{(1/2)}_{n_2}(\tilde{Q}) \equiv
\sqrt{r_0} \prod_{j=1}^{n_2}(\tilde{Q}-\rho_j) \ ,\hspace{3em}
P^{(2)}_{n_2}(\tilde{Q}^2) =
P^{(1/2)}_{n_2}(\tilde{Q})^\dagger P^{(1/2)}_{n_2}(\tilde{Q}) \ .
\ee
 Using this form, the noise vector $\eta$ necessary in the noisy
 correction step can be generated from the gaussian vector
 $\eta^\prime$ according to
\be\label{eq2.1_10}
\eta=P^{(1/2)}_{n_2}(\tilde{Q})^{-1}\eta^\prime \ ,
\ee
 where $P^{(1/2)}_{n_2}(\tilde{Q})^{-1}$ can be obtained as
\be\label{eq2.1_11}
P^{(1/2)}_{n_2}(\tilde{Q})^{-1} =
\frac{P^{(1/2)}_{n_2}(\tilde{Q})^\dagger}
{P^{(2)}_{n_2}(\tilde{Q}^2)} \simeq
P_{n_3}(\tilde{Q}^2) P^{(1/2)}_{n_2}(\tilde{Q})^\dagger \ .
\ee
 In the last step $P_{n_3}$ denotes a polynomial approximation for
 the inverse of $P^{(2)}_{n_2}$ on the interval $[\epsilon,\lambda]$.
 Note that this last approximation can also be replaced by an iterative
 inversion of $P^{(2)}_{n_2}(\tilde{Q}^2)$.
 However, tests showed that the inversion by a least-squares optimized
 polynomial approximation is much faster because, for a given precision,
 less matrix multiplications have to be performed.

 In most of our Monte Carlo computations presented in this paper we
 used the second form in (\ref{eq2.1_10})-(\ref{eq2.1_11}).
 The first form could, however, be used as well.
 In fact, for very high orders $n_2$ or on a 32-bit computer the first
 scheme would be better from the point of view of rounding errors.
 The reason is that in the second scheme for the evaluation of
 $P^{(1/2)}_{n_2}(\tilde{Q})$ we have to use the product form in terms
 of the roots $\rho_j$ in (\ref{eq2.1_9}).
 Even using the optimized ordering of roots defined in
 \cite{TWO-STEP,POLYNOM}, this is numerically less stable than the
 recursive evaluation according to (\ref{apA_4}), (\ref{apA_10}).
 If one uses the first scheme both $P^{(2)}_{n_2}$ in (\ref{eq2.1_4})
 and $R_{n_3}$ in (\ref{eq2.1_6})-(\ref{eq2.1_7}) can be evaluated
 recursively.
 Nevertheless, on our 64-bit machine both methods worked well and we
 have chosen to apply (\ref{eq2.1_11}) where the determination of
 the least-squares optimized polynomials is somewhat simpler.

 The global accept-reject step for the gauge field has been performed
 in our simulations after full sweeps over the gauge field links.
 The order $n_1$ of the first polynomial $P^{(1)}_{n_1}$ has been chosen
 such that the average acceptance probability of the noisy correction
 was near 90\%.
 In principle one can decrease $n_1$ and/or increase the acceptance
 probability by updating only some subsets of the links before the
 accept-reject step.
 This might be useful on lattices larger than our largest lattice
 $12^3 \cdot 24$, but in our case we could proceed with full gauge
 sweeps  and this seemed to be advantageous from the point of view of
 autocorrelations.

%%%%%%%%%%%%%%%%%%%%%%%%%%%%%%%%%%%%%%%%%%%%%%%%%%%%%%%%%%%%%%%%%%%%%%%%
\subsection{Measurement correction: reweighting}\label{sec2.2}

 The multi-bosonic algorithms become exact only in the limit of
 infinitely high polynomial orders: $n\to\infty$ in (\ref{eq2_2}) or, in
 the two-step approximation scheme,  $n_2\to\infty$ in (\ref{eq2_6}).
 Instead of investigating the dependence on the polynomial order by
 performing several simulations, it is practically better to fix some
 high order for the simulation and perform another correction in the
 ``measurement'' of expectation values by still finer polynomials.
 This is done by {\em reweighting} the configurations in the measurement
 of different quantities.
 In case of $N_f=2$ flavours this kind of reweighting has been used
 in \cite{FREJAN} within the polynomial hybrid Monte Carlo scheme.
 As remarked above, $N_f=2$ is special because the reweighting can be
 performed by an iterative inversion.
 The general case can, however, also be treated by a further polynomial
 approximation.

\begin{figure}[htb]
\begin{center}
\epsfig{file=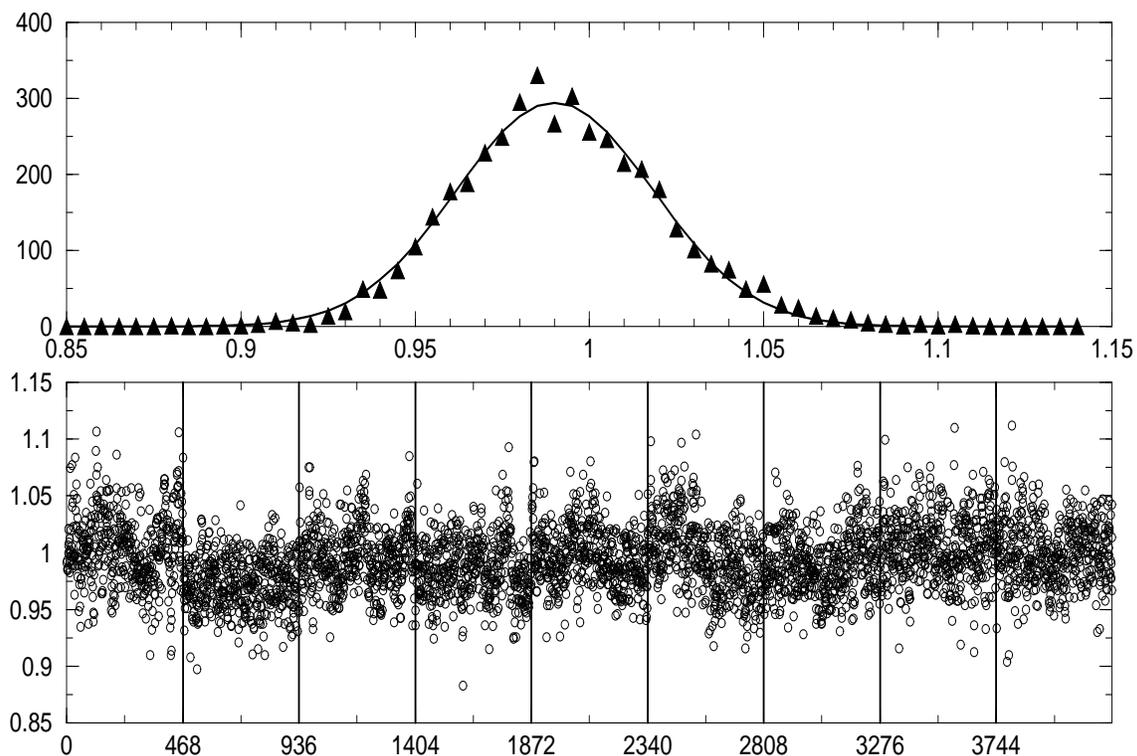,
        width=15cm,height=10cm}
\end{center}
\vspace{-1.0em}
\begin{center}
\parbox{15cm}{\caption{\label{corrdistr_fig}\em
 The distribution of the correction factors in nine independent
 (parallel) sequences of configurations on $12^3 \cdot 24$ lattice at
 $\beta=2.3, K=0.1925$.
 The considered configurations are separated by 50 updating cycles.
 The upper part shows the distribution and a Gaussian fit.
 In the lower part the independent lattices are separated by vertical
 lines.
}}
\end{center}
\end{figure}

 The measurement correction for general $N_f$ has been introduced in
 \cite{BOULDER}.
 It is based on a polynomial approximation $P^{(4)}_{n_4}$ which
 satisfies
\be\label{eq2.2_1}
\lim_{n_4 \to \infty} P^{(1)}_{n_1}(x)P^{(2)}_{n_2}(x)P^{(4)}_{n_4}(x) =
x^{-N_f/2} \ , \hspace{3em}
x \in [\epsilon^\prime,\lambda] \ .
\ee
 The interval $[\epsilon^\prime,\lambda]$ can be chosen, for instance,
 such that $\epsilon^\prime=0,\lambda=\lambda_{max}$, where
 $\lambda_{max}$ is an absolute upper bound of the eigenvalues of
 $Q^\dagger Q=\tilde{Q}^2$.
 In this case the limit $n_4\to\infty$ is exact on an arbitrary gauge
 configuration.
 For the evaluation of $P^{(4)}_{n_4}$ one can use $n_4$-independent
 recursive relations (see appendix \ref{appenA}), which can be stopped
 by observing the convergence of the result.
 After reweighting the expectation value of a quantity $A$ is given by
\be\label{eq2.2_2}
\langle A \rangle = \frac{
\langle A \exp{\{\eta^\dagger[1-P^{(4)}_{n_4}(Q^\dagger Q)]\eta\}}
\rangle_{U,\eta}}
{\langle  \exp{\{\eta^\dagger[1-P^{(4)}_{n_4}(Q^\dagger Q)]\eta\}}
\rangle_{U,\eta}} \ ,
\ee
 where $\eta$ is a simple Gaussian noise like $\eta^\prime$ in
 (\ref{eq2.1_5}).
 Here $\langle\ldots\rangle_{U,\eta}$ denotes an expectation value
 on the gauge field sequence, which is obtained in the two-step process
 described in the previous subsection, and on a sequence of independent
 $\eta$'s.
 The expectation value with respect to the $\eta$-sequence can be
 considered as a Monte Carlo updating process with the trivial action
 $S_\eta \equiv \eta^\dagger\eta$.
 The length of the $\eta$-sequence on a fixed gauge configuration can
 be, in principle, arbitrarily chosen.
 In praxis it has to be optimized for obtaining the smallest possible
 errors.
 If the second polynomial gives a good approximation the correction
 factors do not practically change the expectation values.
 A typical example is shown in figure~\ref{corrdistr_fig}.
 In such cases the measurement correction is good for the confirmation
 of the results.

 The application of the measurement correction is most important for
 quantities which are sensitive for small eigenvalues of the fermion
 matrix $Q^\dagger Q$.
 The polynomial approximations are worst near $x=0$ where the function
 $x^{-N_f/2}$ diverges.
 In the exact effective gauge action, including the fermion determinant,
 the configuration with small eigenvalues $\Lambda$ are suppressed by
 $\Lambda^{N_f/2}$.
 The polynomials at finite order are not able to provide such a strong
 suppression, therefore in the updating sequence of the gauge fields
 there are more configurations with small eigenvalues than needed.
 The {\em exceptional configurations} with exceptionally small
 eigenvalues have to be suppressed by the reweighting.
 This can be achieved by choosing $\epsilon^\prime=0$ and a high enough
 order $n_4$.
 It is also possible to take some non-zero $\epsilon^\prime$ and
 determine the eigenvalues below it exactly.
 Each eigenvalue $\Lambda < \epsilon^\prime$ is taken into account by an
 additional reweighting factor
 $\Lambda^{N_f/2}P^{(1)}_{n_1}(\Lambda)P^{(2)}_{n_2}(\Lambda)$.
 The stochastic correction in (\ref{eq2.2_2}) is then restricted to the
 subspace orthogonal to these eigenvectors.
 Instead of $\epsilon^\prime > 0$ one can also keep $\epsilon^\prime=0$
 and project out a fixed number of smallest eigenvalues.
 Since the control of the smallest eigenvalues of the fermion matrix
 is an essential part of our simulations, a short summary of the
 numerical methods to obtain them is included in appendix \ref{appenB}.

 Let us note that, in principle, it would be enough to perform just a
 single kind of correction.
 But to omit the reweighting does not pay because it is much more
 comfortable to investigate the (small) effects of different $n_4$
 values on the expectation values than to perform several simulations
 with increasing values of $n_2$.
 Without the updating correction the whole correction could be done by
 reweighting in the measurements.
 However, in practice this would not work either.
 The reason is that a first polynomial with relatively low order does
 not sufficiently suppress the exceptional configurations.
 As a consequence, the reweighting factors would become too small and
 would reduce the effective statistics considerably.
 In addition, the very small eigenvalues are changing slowly in the
 update and this would imply longer autocorrelations.

\begin{figure}[htb]
\begin{center}
\epsfig{file=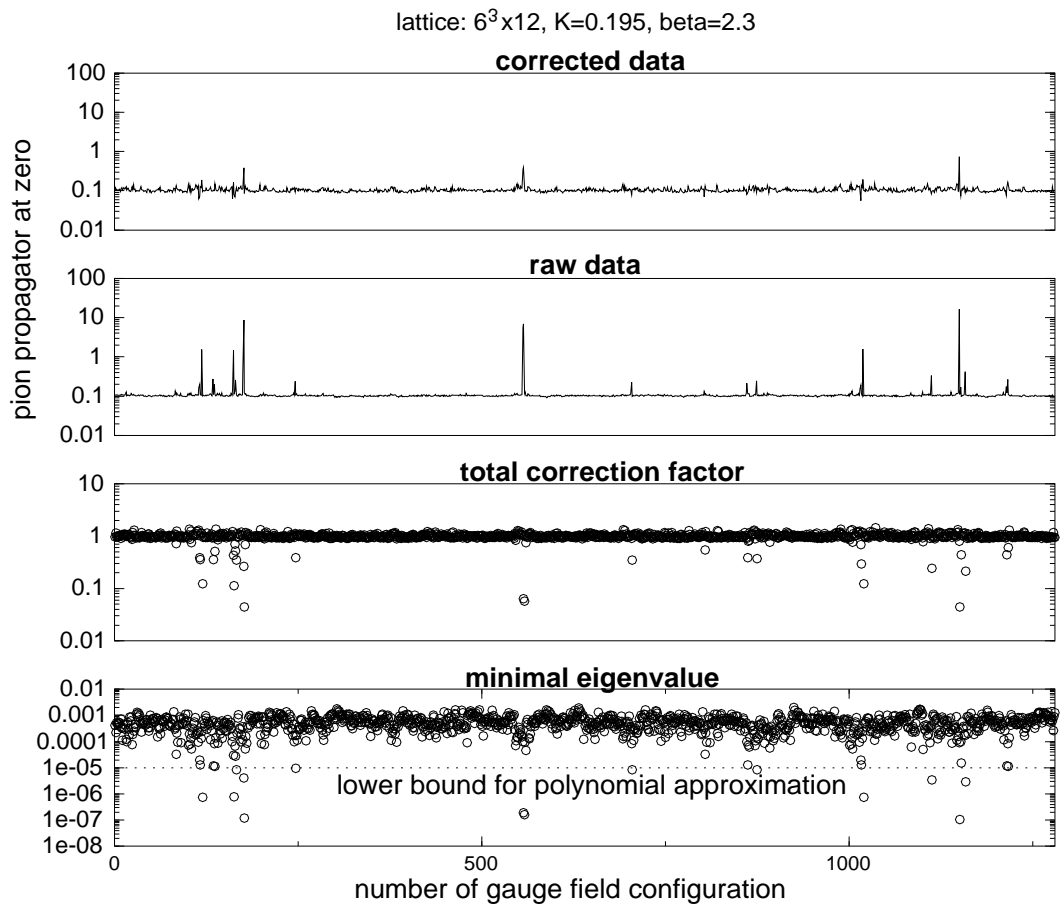,
        width=14cm,height=12cm}
\end{center}
\vspace{-1.0em}
\begin{center}
\parbox{15cm}{\caption{\label{corr_fig}\em
 The measurement correction for the a-pion propagator at zero distance.
 The exceptional configurations with small eigenvalues contribute
 strongly to the raw data.
 After correction these contributions are still important but of
 normal size.
}}
\end{center}
\end{figure}

 A moderate surplus of gauge configurations with small eigenvalues may,
 however, be advantageous because it allows for a better sampling of
 such configurations and enhances the tunneling among sectors with
 different topological charges.
 For small fermion masses on large physical volumes this is expected to
 be more important than the prize one has to pay for it by reweighting,
 provided that the reweighting has only a moderate effect.
 The effect of a better sampling of configurations which small
 eigenvalues can be best illustrated by the distribution of quantities
 which diverge for zero eigenvalues.
 An example on $6^3 \cdot 12$ lattice at $\beta=2.3,K=0.195$ is shown
 in figure~\ref{corr_fig}.

%%%%%%%%%%%%%%%%%%%%%%%%%%%%%%%%%%%%%%%%%%%%%%%%%%%%%%%%%%%%%%%%%%%%%%%%
\subsection{The Pfaffian and its sign}\label{sec2.3}

 The Pfaffian resulting from the Grassmannian path integrals for
 Majorana fermions (\ref{eq1.1_7}) is an object similar to a determinant
 but less often used \cite{PFAFFIAN}.
 As shown by (\ref{eq1.1_8}), ${\rm Pf}(M)$ is a polynomial of the
 matrix elements of the $2N$-dimensional antisymmetric matrix $M=-M^T$.
 Basic relations are
\be\label{eq2.3_1}
M = P^T J P ,\hspace{2em}  {\rm Pf}(M)=\det(P) \ ,
\ee
 where $J$ is a block-diagonal matrix containing on the diagonal
 $2\otimes2$ blocks equal to $\epsilon=i\sigma_2$ and otherwise zeros.
 From this eq.~(\ref{eq1.1_9}) immediately follows.

 The form of $M$ in (\ref{eq2.3_1}) can be achieved by a procedure
 analogous to the Gram-Schmidt orthogonalization and, by construction,
 $P$ is a triangular matrix.
 In order to see this, let us introduce the notation
\be\label{eq2.3_2}
( u M v ) \equiv
\sum_{\alpha,\beta=1}^{2N} u_\alpha M_{\alpha\beta} v_\beta =
(v M^T u)
\ee
 and denote the orthonormal basis vectors by
 $\{ e_\alpha,\; \alpha=1,2,\ldots,2N \}$.
 We are looking for a new basis $\{ a_j,b_k,\; j,k=1,2,\ldots,N \}$
 obtained by
\be\label{eq2.3_3}
a_j \equiv P e_{2j-1} = \sum_\alpha e_\alpha\; (e_\alpha P e_{2j-1}) \ ,
\hspace{3em}
b_k \equiv P e_{2k} = \sum_\alpha e_\alpha\; (e_\alpha P e_{2k})
\ee
 such that the matrix elements on it are given by
\be\label{eq2.3_4}
(a_j M a_k) = 0 \ , \hspace{2em}
(b_j M b_k) = 0 \ , \hspace{2em}
(b_k M a_j) = -(a_j M b_k) = \delta_{jk} \ .
\ee
 The construction is started by defining
\be\label{eq2.3_5}
a_1 = e_1 \ , \hspace{3em} b_1 = \frac{e_2}{M_{21}} \ .
\ee
 (If $M_{21}$ is zero one has to rearrange the ordering of the original
 basis to achieve $M_{21} \ne 0$.)
 In the next step $e_l,\; l=3,4,\ldots,2N$ is replaced by
\be\label{eq2.3_6}
e^\prime_{l-2} \equiv e_l - a_1\, (b_1 M e_l) + b_1\, (a_1 M e_l) \ ,
\ee
 which satisfy
\be\label{eq2.3_7}
(e^\prime_{l-2} M a_1) = (e^\prime_{l-2} M b_1) = 0 \ .
\ee
 With this the required form in (\ref{eq2.3_4}) is achieved for
 $a_1$ and $b_1$ and the corresponding matrix elements of $P$ in
 (\ref{eq2.3_3}), which are necessary for $a_1$ and $b_1$, are
 determined.
 To proceed one has to return to (\ref{eq2.3_5}) with
 $\{ e_\alpha,\; \alpha=1,2,\ldots,2N \}$ replaced by
 $\{ e^\prime_\alpha,\; \alpha=1,2,\ldots,2N-2 \}$ and obtain the next
 $(a,b)$-pair, until the whole space is exhausted.
 This gives a numerical procedure for the computation of $P$ and the
 determinant of $P$ gives, according to (\ref{eq2.3_1}), the Pfaffian
 ${\rm Pf}(M)$.
 Since $P$ is (lower-) triangular, the calculation of $\det P$ is, of
 course, trivial.

 This procedure can be used for a numerical determination of the
 Pfaffian on small lattices \cite{BOULDER}.
 On lattices larger than, say, $4^3 \cdot 8$ the computation becomes
 cumbersome due to the large storage requirements.
 This is because one has to store a full $\Omega \otimes \Omega$ matrix,
 with $\Omega$ being the number of lattice points multiplied by the
 number of spinor-colour indices (equal to $4(N_c^2-1)$ for the adjoint
 representation of ${\rm SU}(N_c)$).
 The difficulty of computation is similar to a computation of the
 determinant of $Q$ with $LU$-decomposition.

 Fortunately, in order to obtain the sign of the Pfaffian occurring in
 the measurement reweighting formula (\ref{eq1.1_10}), one can proceed
 without a full calculation of the value of the Pfaffian.
 The method is to monitor the sign changes of ${\rm Pf}(M)$ as a
 function of the hopping parameter $K$.
 Since at $K=0$ we have ${\rm Pf}(M)=1$, the number of sign changes
 between $K=0$ and the actual value of $K$, where the dynamical fermion
 simulation is performed, determines the sign of ${\rm Pf}(M)$.
 The sign changes of ${\rm Pf}(M)$ can be determined by the flow of
 the eigenvalues of $\tilde{Q}$ through zero.
 As remarked already in the discussion before (\ref{eq1.1_10}), the
 fermion matrix for the gluino $\tilde{Q}$ has doubly degenerate real
 eigenvalues therefore
\be\label{eq2.3_8}
\det M = \det \tilde{Q} = \prod_{i=1}^{\Omega/2} \tilde{\lambda}_i^2 \ ,
\ee
 where $\tilde{\lambda}_i$ denotes the eigenvalues of $\tilde{Q}$.
 This implies
\be\label{eq2.3_9}
|{\rm Pf}(M)| =  \prod_{i=1}^{\Omega/2} |\tilde{\lambda}_i| \ ,
\hspace{2em} \Longrightarrow \hspace{2em}
{\rm Pf}(M) =  \prod_{i=1}^{\Omega/2} \tilde{\lambda}_i \ .
\ee
 The first equality trivially follows from (\ref{eq1.1_9}).
 The second one is the consequence of the fact that ${\rm Pf}(M)$ is
 a polynomial in $K$ which cannot have discontinuities in any of its
 derivatives.
 Therefore if, as a function of $K$, an eigenvalue $\tilde{\lambda}_i$
 (or any odd number of eigenvalues) changes sign the sign of
 ${\rm Pf}(M)$ has to change, too.
 We tested the sign of the Pfaffian in our Monte Carlo simulations by
 this {\em spectral flow} method.

\begin{figure}[htb]
\begin{center}
\epsfig{file=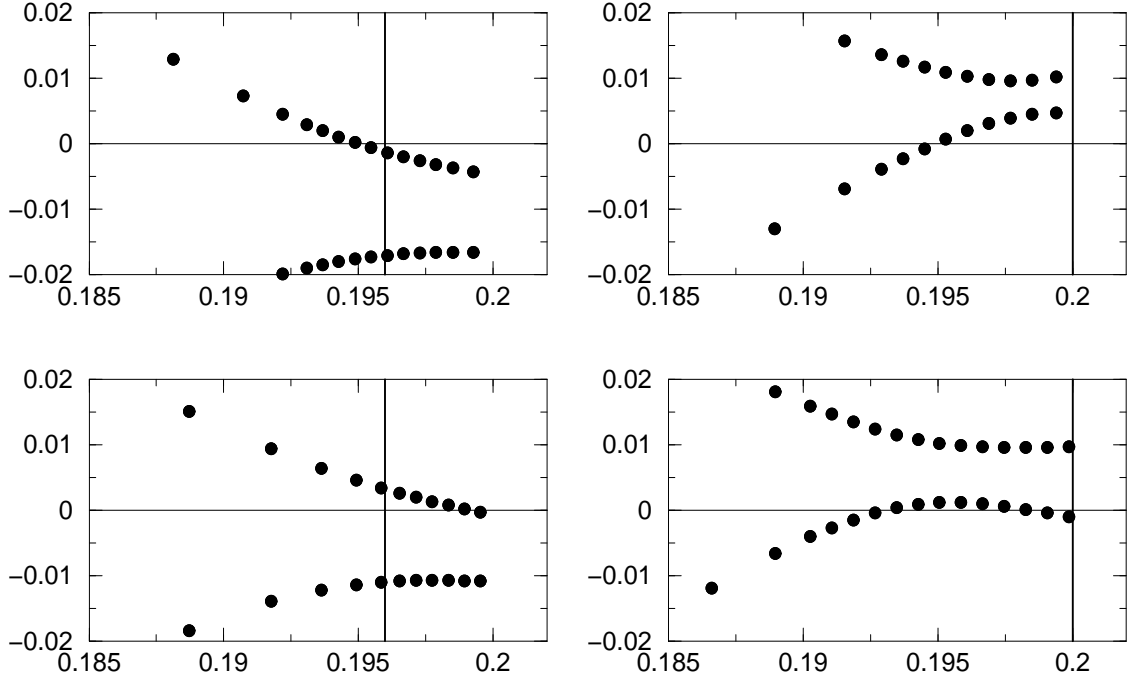,
        width=15cm,height=9cm}
\end{center}
\vspace{-1.5em}
\begin{center}
\parbox{15cm}{\caption{\label{spflow1_fig}\em
 The spectral flow of the hermitean fermion matrix $\tilde{Q}$ for some
 specific configurations on $6^3 \cdot 12$ lattice at $\beta=2.3$.
 The value of $K$ in the simulation is displayed by a vertical line.
}}
\end{center}
\end{figure}

\begin{figure}[htb]
\begin{center}
\epsfig{file=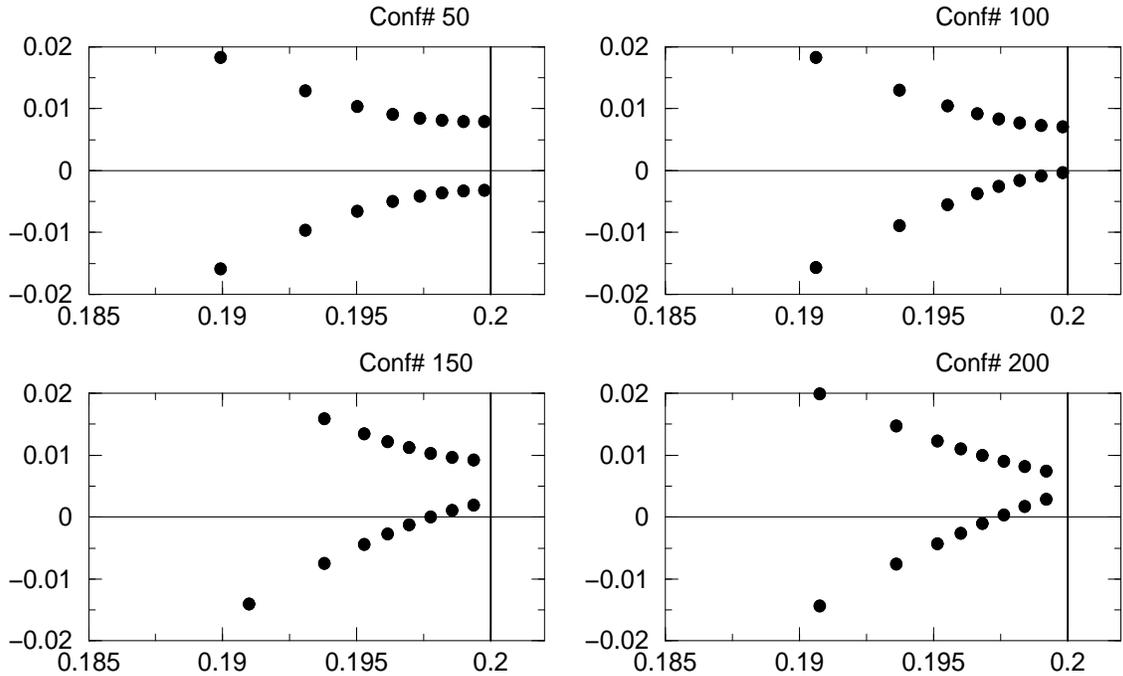,
        width=15cm,height=9cm}
\end{center}
\vspace{-1.5em}
\begin{center}
\parbox{15cm}{\caption{\label{spflow2_fig}\em
 The spectral flow of the hermitean fermion matrix
 $\tilde{Q}$ for some configurations separated by 50 updating cycles on
 $6^3 \cdot 12$ lattice at $\beta=2.3;\; K=0.2$.
}}
\end{center}
\end{figure}

 As a representative example, let us consider the Monte Carlo runs on
 $6^3 \cdot 12$ lattice for $K=0.19,\; 0.196,\; 0.20$.
 The number of gauge configurations with negative Pfaffian in some
 representative subsets of the measured gauge configurations is
 given in table~\ref{pfaffsign_tab}.
 The flow of the lowest eigenvalues with the hopping parameter $K_v$
 is shown in some examples in figures~\ref{spflow1_fig},
 \ref{spflow2_fig}.
 The conclusion is that the probability of negative Pfaffians at most
 parameter values is negligible.
 Only at the largest hopping parameter, which corresponds to a
 negative gluino mass beyond the chiral phase transition
 \cite{PHASETRANS}, there is a somewhat larger fraction with negative
 Pfaffians but their effect on the averages is still smaller than the
 statistical errors.
 Therefore taking the absolute value of the Pfaffian, as in
 eq.~(\ref{eq1.1_1}), gives in the physically interesting points a
 very good approximation.
%%%%%%%%%%%%%%%%%%%%%%%%%%%%%%%%%%%%%%%%%%%%%%%%%%%%%%%%%%%%%%%%%%%%%%
\vspace*{-1.0em}
\begin{table}[ht]
\begin{center}
\parbox{15cm}{\caption{\label{pfaffsign_tab}\em
 The fraction of Pfaffians with negative sign at $\beta=2.3$ on
 $6^3 \cdot 12$ lattice for different hopping parameters $K$.
}}
\end{center}
\begin{center}
\begin{tabular}{| l || r | c | l ||}
\hline
K  &  \# configs.  &  \# of $Pf(M)<0$  &  fraction \\
\hline
\hline
0.19  &  3840 (60x64)  &  0   & $<$ 0.0003 \\ \hline
0.196 &  5248 (82x64)  & 14   &     0.0027 \\ \hline
0.2   &  2304 (36x64)  & 69   &     0.03   \\ \hline
\end{tabular}
\end{center}
\end{table}

%%%%%%%%%%%%%%%%%%%%%%%%%%%%%%%%%%%%%%%%%%%%%%%%%%%%%%%%%%%%%%%%%%%%%%%%
\subsection{Preconditioning}\label{sec2.4}

 The difficulty of numerical simulations increases with the {\em
 condition number} $\lambda/\epsilon$ characterizing the eigenvalue
 spectrum of fermion matrices on typical gauge field configurations.
 As it is well known, one can decrease the condition number by
 {\em preconditioning}.
 Even-odd preconditioning in multi-bosonic algorithms have been
 introduced in \cite{JEGERL}.
 This turned out to be very useful in our simulations.

 For even-odd preconditioning the hermitean fermion matrix $\tilde{Q}$
 is decomposed in subspaces containing the odd, respectively, even
 points of the lattice as
\be\label{eq2.4_1}
\tilde{Q} = \gamma_5 Q = \left(
\begin{array}{cc}
\gamma_5  &  -K\gamma_5 M_{oe}  \\  -K\gamma_5 M_{eo}  &  \gamma_5
\end{array} \right) \ .
\ee
 For the fermion determinant we have
\be\label{eq2.4_2}
\det\tilde{Q} = \det\hat{Q} \ ,
\hspace{2em} {\rm with} \hspace{2em}
\hat{Q} \equiv \gamma_5 - K^2\gamma_5 M_{oe}M_{eo} \ .
\ee
 The matrix {$\hat{Q}^2$} has a smaller condition number than
 {$\tilde{Q}^2$}.

\begin{figure}[htb]
\begin{center}
\epsfig{file=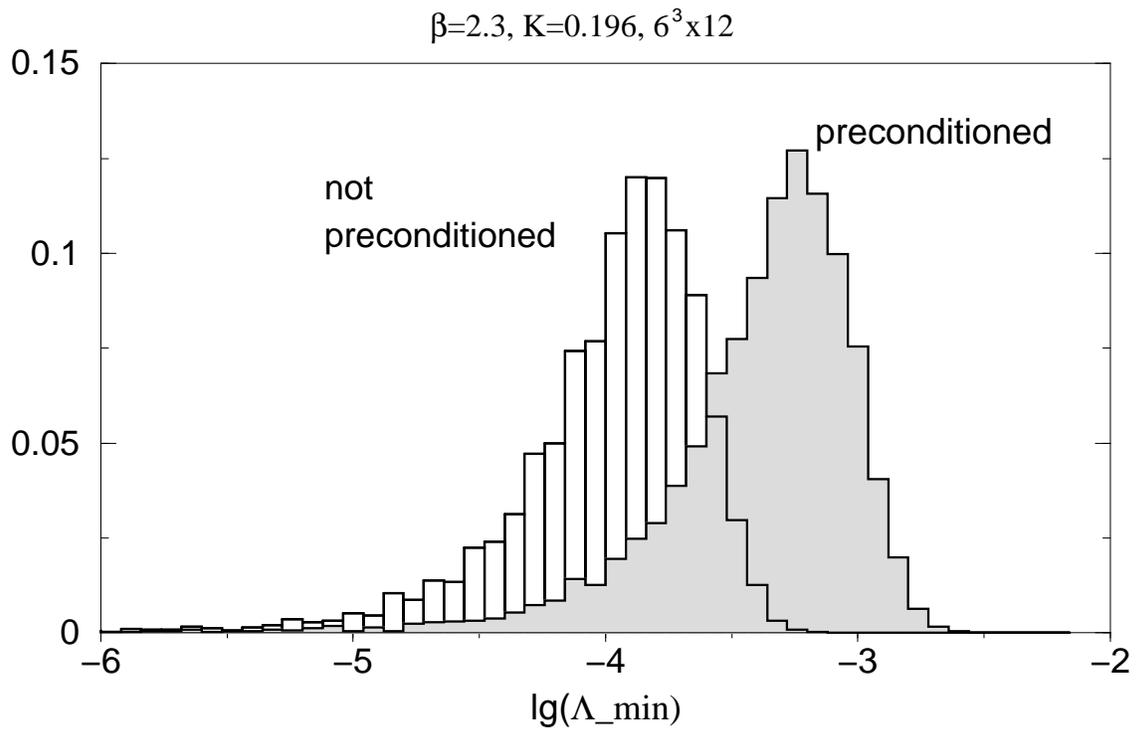,
        width=15cm,height=10cm}
\end{center}
\vspace*{-1.5em}
\begin{center}
\parbox{15cm}{\caption{\label{precond_fig}\em
 The distribution of the smallest eigenvalues of the squared
 preconditioned fermion matrix $\hat{Q}^2$ versus the non-preconditioned
 one $\tilde{Q}^2$ on a $6^3 \cdot 12$ lattice at $\beta=2.3,K=0.196$.
 }}
\end{center}
\end{figure}

 The condition number and its fluctuations on different gauge
 configurations are dominated by the minimal eigenvalue.
 An example of a comparison of the fluctuations of the lowest eigenvalue
 of $\hat{Q}^2$ and $\tilde{Q}^2$ is shown in figure~\ref{precond_fig}.
 As one sees, in this case the mean of the smallest eigenvalue becomes
 about a factor 4 larger due to preconditioning.
 At the same time the largest eigenvalue becomes smaller, therefore the
 average condition number becomes about a factor 5 smaller.

%%%%%%%%%%%%%%%%%%%%%%%%%%%%%%%%%%%%%%%%%%%%%%%%%%%%%%%%%%%%%%%%%%%%%%%%
\subsection{Parameter choice and autocorrelations}\label{sec2.5}

 The two-step multi-bosonic algorithm has several algorithmic parameters
 which can be tuned to achieve optimal performance.
 In fact, our experience shows that this tuning can bring substantial
 gain in efficiency.

\begin{figure}[htb]
\begin{center}
\epsfig{file=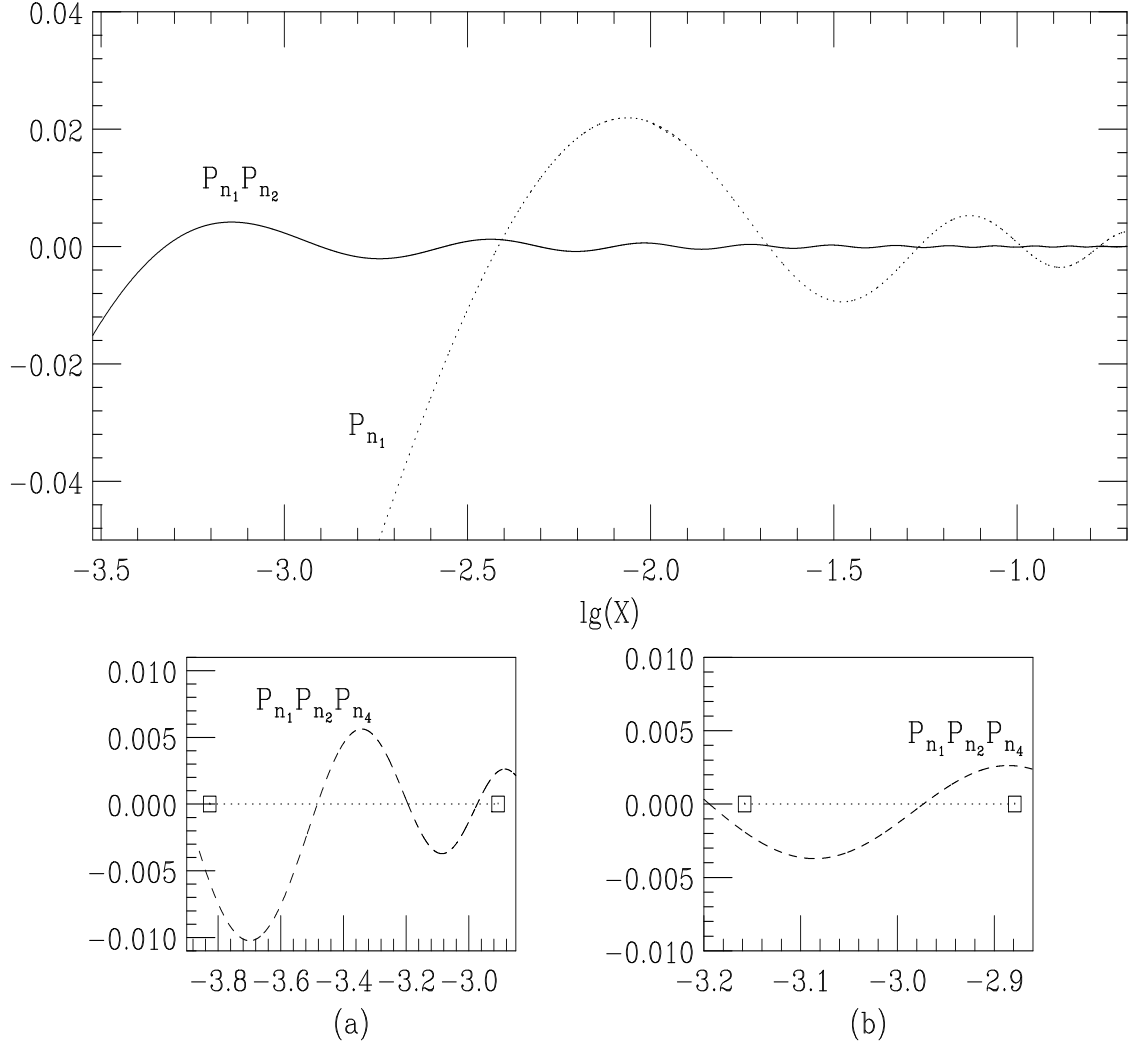,
        width=14cm,height=15cm,angle=90}
\end{center}
\vspace*{-2.0em}
\begin{center}
\parbox{15cm}{\caption{\label{p1p2p4_fig}\em
 Relative deviation of the successive polynomial approximations of
 $x^{-1/4}$ in the range of eigenvalues corresponding to our simulations
 in the $12^3 \cdot 24$ lattice at $K=0.1925$.
 (For parameters see table~\protect{\ref{runs_tab}}.)
}}
\end{center}
\end{figure}

 {\bf Polynomial degrees:}
 In order to fix the polynomial degrees $n_{1\ldots4}$, in practice one
 performs trial runs using increasing values.
 At the same time, by observing the range of eigenvalues, one also
 obtains the interval $[\epsilon,\lambda]$.
 The final value of $n_1$ is fixed by ensuring a high acceptance rate,
 around 90\%, in the update correction step.
 $n_2$ has to be large enough to keep the measurement correction small
 in important physical quantities.
 The final precision of the updating is set by $n_4$, therefore the
 choice of $n_1$ and $n_2$ does not influence the expectation values.
 For showing a typical example, in the upper part of
 figure~\ref{p1p2p4_fig} the polynomial approximation $P_{n_1}^{(1)}$ and
 the product $P_{n_1}^{(1)} P_{n_2}^{(2)}$ are plotted in the interval
 $[\epsilon, \lambda]$.
 The product $P_{n_1}^{(1)} P_{n_2}^{(2)} P_{n_4}^{(4)}$ is displayed in
 the lower part of the figure.
 To the left (label a) the interval covers the range of the fluctuating
 smallest eigenvalue, whereas to the right (label b) the function is
 shown in the range of fluctuations of the last small eigenvalue which
 was determined explicitly.
 (In this case the correction factors were calculated from the eight
 smallest eigenvalues exactly and in the orthogonal subspace
 stochastically.)

 To fix the degree of the third polynomial, $n_3$, we consider the
 probability $p_1$ of the system to jump between two identical
 configurations.
 In the limit $n_3 \rightarrow \infty$ this probability tends obviously
 to 1.
 In practice $n_3$ is increased till we get $p_1 \approx 0.99$, which is
 acceptable from the algorithm precision point of view, as one is
 convinced by comparing the expectation values.
 The choice of $n_4$ has to be tested, in principle, by observing its
 effect on the expectation values.
 Usually it is possible to choose already $n_2$ so large that the
 measurement corrections with a substantially higher $n_4$ are
 negligible compared with the statistical errors.

 The parameters of the numerical simulations at $\beta=2.3$ are
 summarized in table~\ref{runs_tab}.
 The runs with an asterisk had periodic boundary conditions for the
 gluino in the time direction $T$, the rest antiperiodic.
 $K$ is the hopping parameter and $[\epsilon,\lambda]$ is the interval
 of approximation for the first three polynomials of orders
 $n_{1,2,3}$, respectively.
 The fourth polynomial of order $n_4$ is defined on $[0,\lambda]$.
 In the last two columns the number of performed updating cycles,
 respectively, the number of parallelly updated lattices ($N_{lat}$) are
 given.
%%%%%%%%%%%%%%%%%%%%%%%%%%%%%%%%%%%%%%%%%%%%%%%%%%%%%%%%%%%%%%%%%%%%%%%%
\begin{table}[ht]
\begin{center}
\parbox{15cm}{\caption{\label{runs_tab}\em
 Parameters of the numerical simulations at $\beta=2.3$.
 The notations are explained in the text.
}}
\end{center}
\begin{center}
\begin{tabular}{| c | l | l | c | c | c | c | c | r | c |}
\hline
lattice: $L,T$  &
\multicolumn{1}{|c|}{$K$}  &
\multicolumn{1}{|c|}{$\epsilon$}  &
$\lambda$  & $n_1$  &  $n_2$  &  $n_3$  &  $n_4$  &
\multicolumn{1}{|c|}{updates}  &  $N_{lat}$  \\
\hline\hline
6,12$^\ast$  &  0.16  &  0.008  &  3.2  &  8  &  32  &  32  &
  -  &  374400  &  64  \\
\hline
6,12$^\ast$  &  0.17  &  0.008  &  3.2  &  8  &  32  &  32  &
  -  &  332800  &  64  \\
\hline
6,12$^\ast$  &  0.18  &  0.008  &  3.2  &  8  &  32  &  32  &
  -  &  540800  &  64  \\
\hline
6,12$^\ast$  &  0.185 &  0.002  &  3.4  &  12  &  32  &  48  &
  -  &  384000  &  64  \\
\hline
6,12  &  0.185 &  0.002  &  3.4  &  16  &  100  &  150  &
  200  &  550400  &  64  \\
\hline
6,12$^\ast$  &  0.19  &  0.0002  &  3.5  &  16  &  60  &  96  &
  -  &  712800  &  64  \\
\hline
6,12  &  0.19    &  0.0005   &  3.6  &  20  &  112  &  150  &  400  &
1487360  &  64  \\
\hline
6,12$^\ast$  &  0.1925  &  0.00003  &  3.7  &  22  &  66  &  102  &
  400  &  1280000  &  64  \\
\hline
6,12  &  0.1925  &  0.0001   &  3.7  &  22  &  132  &  180  &  400  &
3655680  &  64  \\
\hline
6,12$^\ast$  &  0.195  &  0.00003  &  3.7  &  22  &  66  &  102  &
  400  &  1224000  &  64  \\
\hline
6,12  &  0.195   &  0.00001  &  3.7  &  24  &  200  &  300  &  400  &
460800  &  64  \\
\hline
6,12  &  0.196   &  0.00001  &  3.7  &  24  &  200  &  300  &  400  &
952320  &  64  \\
\hline
6,12  &  0.1975  &  0.000001 &  3.8  &  30  &  300  &  400  &  500  &
506880  &  64  \\
\hline
6,12  &  0.2     &  0.000001 &  3.9  &  30  &  300  &  400  &  500  &
599040  &  64  \\
\hline\hline
8,16  &  0.19    &  0.00065  &  3.55 &  20  &  82   &  112  &  -  &
1038400  &  32  \\
\hline
8,16  &  0.1925  &  0.0001   &  3.6  &  22  &  142  &  190  &  -  &
870400  &  32  \\
\hline\hline
12,24  &  0.1925  &  0.0003 &  3.7  &  32  &  150  &  220  &  400  &
216000  &  9  \\
\hline\hline
\end{tabular}
\end{center}
\end{table}
%%%%%%%%%%%%%%%%%%%%%%%%%%%%%%%%%%%%%%%%%%%%%%%%%%%%%%%%%%%%%%%%%%%%%%%%

{\bf Optimal ordering of the roots:}
 The roots of $P_{n_1}^{(1)}$ have to be always calculated.
 As discussed in section \ref{sec2.1}, depending on the way of doing the
 global accept-reject in updating, sometimes the roots of
 $P_{n_2}^{(2)}$ are also needed.
 Concerning this point a non-trivial question is how to order the roots
 when the representation (\ref{eq2_4}) is used.
 Choosing this order naively leads to overflow and underflow problems
 because the product in (\ref{eq2_4}) involves in general very different
 orders of magnitude.
 A good solution \cite{POLYNOM} is minimizing the maximal ratio of the
 values $x^{\alpha} P_p(x)$ for $x \in [\epsilon,\lambda]$,
 where $P_p(x)$ denotes the partial product under consideration.
 This is in practice achieved by considering a discrete number of points
 in the interval $\{x_1,\ldots,x_N\}$ where $N={\cal O}(n)$.
 This gives in general sufficient numerical stability even for orders
 of many hundreds (see also the tests performed in \cite{MONOMIAL}).

{\bf Autocorrelations:}
 During our simulations autocorrelations of different quantities were
 determined.
 Here we report on the analysis for the $12^3 \cdot 24$ lattice at
 $\beta = 2.3,K = 0.1925$.
 Results for the $6^3 \cdot 12$ and the $8^3 \cdot 16$ lattices can be
 found in \cite{PHASETRANS,SPANDEREN}.
 We considered the short range exponential autocorrelation $\tau_{exp}$
 of three different quantities, namely
\begin{itemize}
\item
the $a$-$\eta^\prime$ propagator
\item
the gluino-glue propagator
\item
the plaquette
\end{itemize}
 In the last case the data has been sufficient to give also an
 estimate of the integrated autocorrelation.

 We started the analysis by calculating the autocorrelation function
 $C(t)$ for all these quantities.
 In case of the $a$-$\eta^\prime$ we calculated the autocorrelation of
 the propagator at time distance $\Delta t=1$, considering every 150'th
 configuration of the updating series.
 This was done separately on each of the lattices run in parallel.
 By averaging over the correlation functions obtained in this way we
 observed that the mean correlation function
 $\bar{C}^{a-\eta^\prime}(t)$ was at $t =1$ already compatible with zero
 ($\bar{C}^{a-\eta^\prime}(1)=0.028(19)$), which lead to the conclusion
 that $\tau^{a-\eta^\prime}_{exp} \leq 150$ updates.

\begin{figure}[htb]
\begin{center}
\epsfig{file=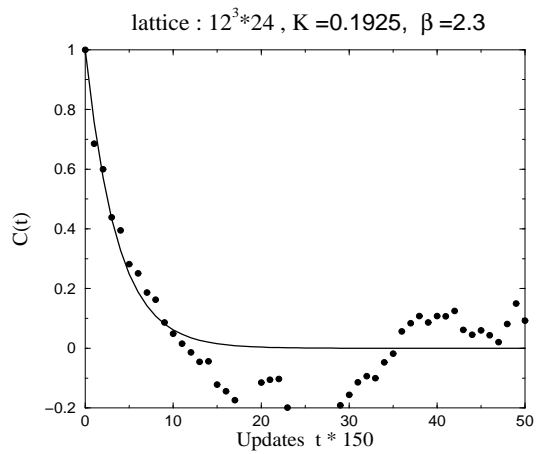,
        width=7cm,height=6cm}
\end{center}
\vspace*{-1.0em}
\begin{center}
\parbox{15cm}{\caption{\label{autocgglu_fig}\em
 Autocorrelation function and exponential fit for the
 gluino-glue propagator on one of the $12^3 \cdot 24$ lattices run in
 parallel at $\beta = 2.3,K = 0.1925$.
}}
\end{center}
\end{figure}

 Estimating the exponential autocorrelation $\tau_{exp}$ of the
 gluino-glue propagator we proceeded similarly as in the
 $a$-$\eta^\prime$ case.
 On all nine lattices that were run in parallel we determined
 independently the autocorrelation of the propagator at time distance
 $\Delta t=1$ on every 150'th configuration of our total history.
 The exponential autocorrelation time was then estimated by
 fitting an exponential of the form $\exp(-t/\tau_{exp})$ to the first
 points of the curve for each lattice.
 A typical autocorrelation function with the exponential fit can be seen
 in figure~\ref{autocgglu_fig}.
 By finally taking the average of $\tau_{exp}$ over all lattices
 we arrived at the result displayed in table~\ref{autocorr_tab}.
 It has to be understood that $\tau_{exp}$ determined in this way
 displays a mode between the true short range exponential and the
 integrated autocorrelation, since only every 150'th sweep has been
 considered.

\begin{figure}[htb]
\begin{flushleft}
\epsfig{file=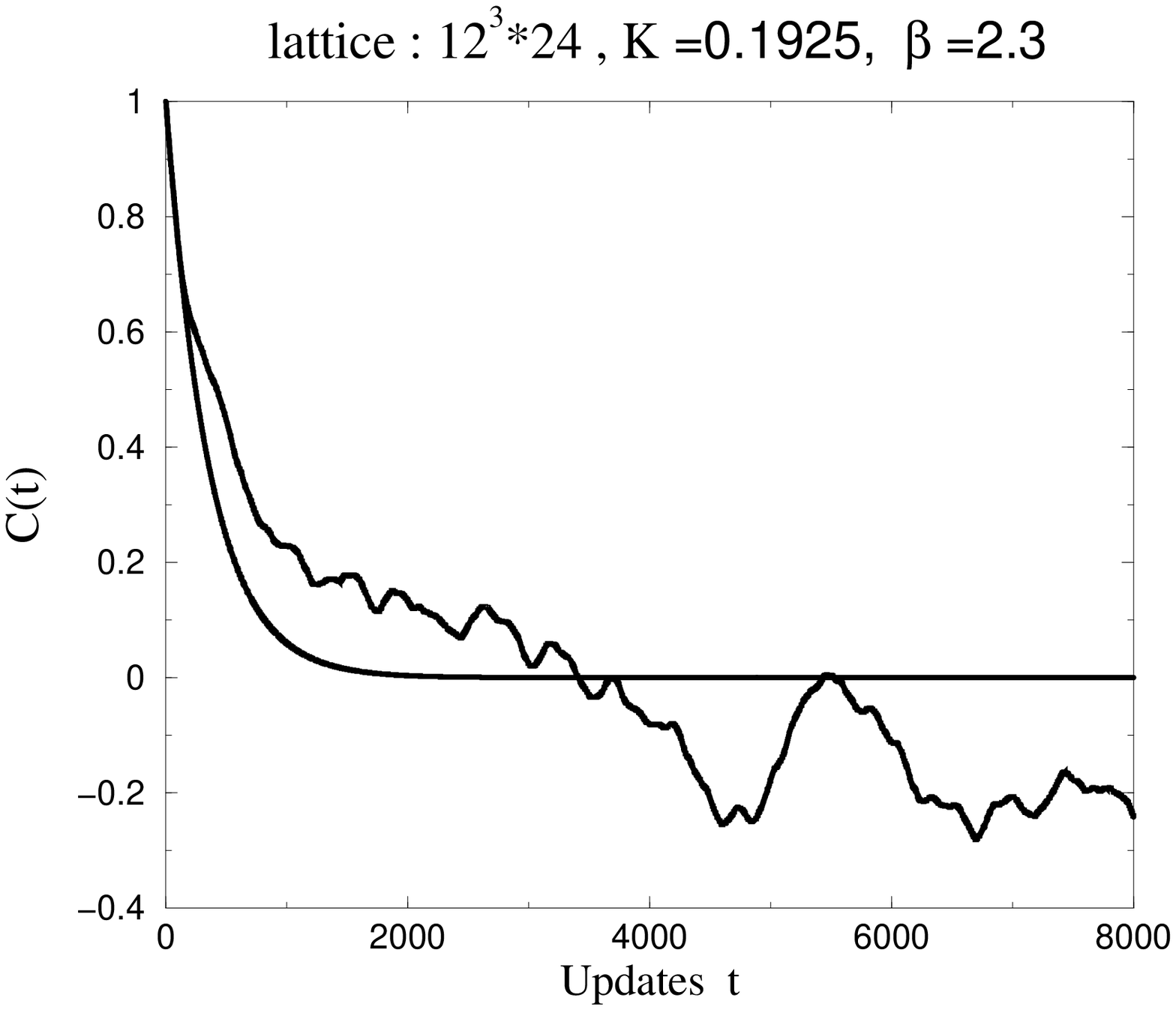,
        width=7cm,height=6cm}
\end{flushleft}
\vspace*{-70mm}
\begin{flushright}
\epsfig{file=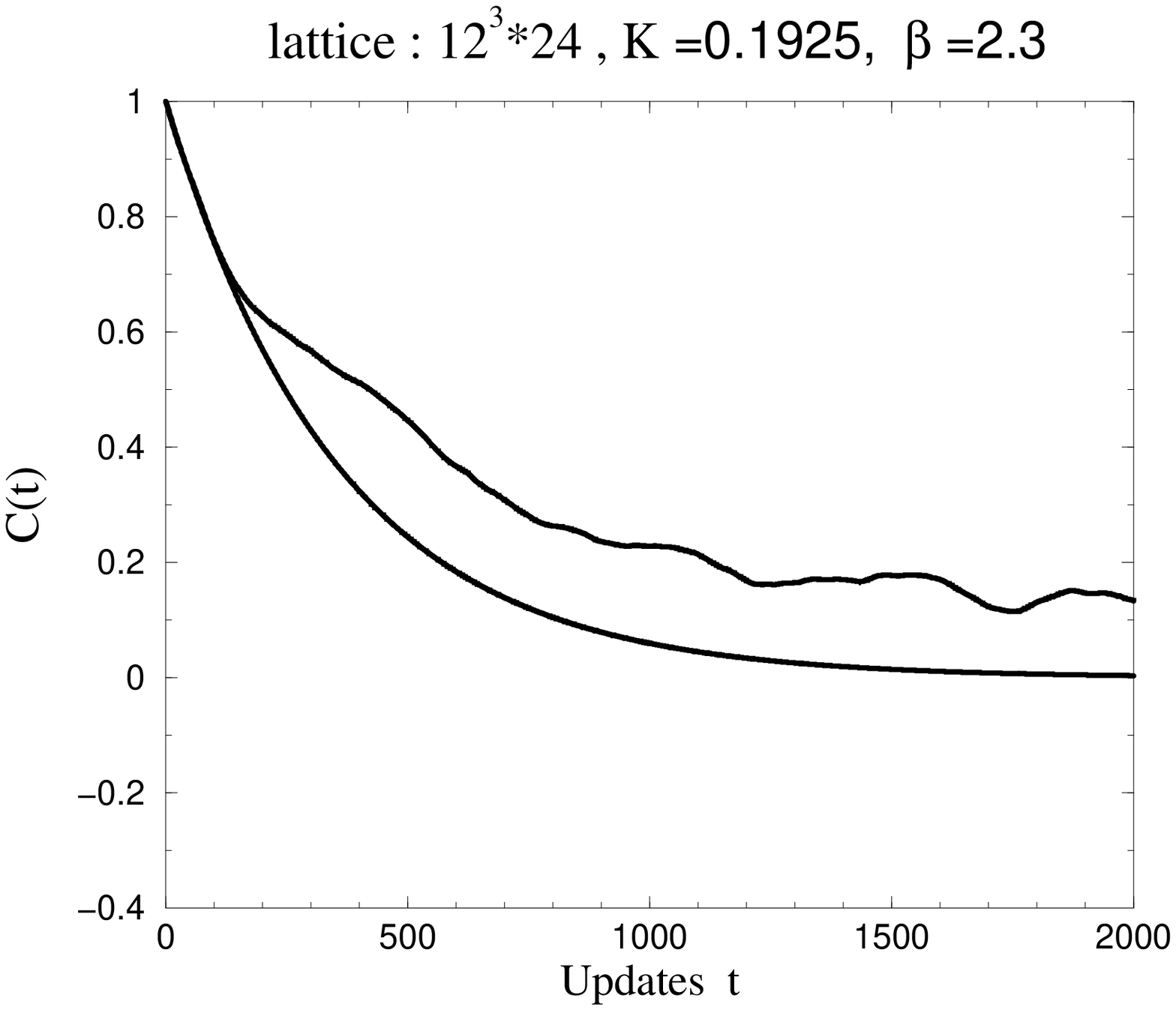,
        width=7cm,height=6cm}
\end{flushright}
\vspace*{-1.0em}
\begin{center}
\parbox{15cm}{\caption{\label{plaqautocorr_fig}\em
 Typical autocorrelation of the plaquette, with the exponential fit.
 The right graph shows the same data in a smaller interval.
}}
\end{center}
\end{figure}

 To estimate the integrated autocorrelation time $\tau_{int}$ of the
 plaquette we proceeded in a different manner.
 On the basis of prior analysis \cite{PHASETRANS,SPANDEREN} we expect
 the order of magnitude of $\tau_{int}$ to be $10^2 \sim 10^3$.
 Since for each lattice we have a total of about $24000$ configurations
 in equilibrium we expect our time history to have a length of at most
 $\sim 100 \tau_{int}$.
 This leads to the conclusion that standard methods to determine
 $\tau_{int}$ \cite{SOKAL,BINNING} are not reliable since they require
 statistics that are at least of the order of several hundred
 $\tau_{int}$.
 Therefore, to estimate the order of magnitude of $\tau_{int}$ we
 proceed as follows.
 For each lattice run in parallel we calculated the autocorrelation
 function $C^{plaq}(t)$ of the plaquette for the complete history of
 $24000$ configurations.
 We fitted an exponential decay to $C^{plaq}(t)$ in a small interval
 (typically $[0,300]$) at the beginning where the fastest decay mode
 should be dominant.
 For longer distance the exponential typically decayed faster than
 $C^{plaq}(t)$.
 This expected behaviour could usually be observed up to a point
 $\hat{t}$ where $C^{plaq}(t)$ started to be dominated by its noise.
 We then calculated
\be\label{eq2.5_1}
\sum_{t=1}^{\hat{t}} C^{plaq}(t)
\ee
 and took this value as an estimate of $\tau_{int}$.
 We expect this procedure only to lead to an order of magnitude estimate
 for the integrated autocorrelation.
 The typical behaviour for the autocorrelation function of the plaquette
 together with the exponential fit can be observed in
 figure~\ref{plaqautocorr_fig}.
 In this example the cutoff $\hat{t}$ has been chosen at
 about $\hat{t} \sim 3500$ updates  since at this point $C^{plaq}(t)$
 is clearly dominated by its noise.
 The final result for  $\tau_{exp}$ and $\tau_{int}$ have been obtained
 by averaging over all nine lattices run in parallel, and can be found
 in table~\ref{autocorr_tab}.
%%%%%%%%%%%%%%%%%%%%%%%%%%%%%%%%%%%%%%%%%%%%%%%%%%%%%%%%%%%%%%%%%%%%%%
\begin{table}[ht]
\begin{center}
\parbox{15cm}{\caption{\label{autocorr_tab}\em
 Autocorrelation and integrated autocorrelation of the propagators and
 the plaquette on $12^3 \cdot 24$ lattice at $\beta=2.3$ and
 $K=0.1925$.
}}
\end{center}
\begin{center}
\begin{tabular}{| l | l | l |}
\hline
\multicolumn{1}{|l|}{            }  &
\multicolumn{1}{|c|}{$\tau_{exp}$}  &
\multicolumn{1}{|c|}{$\tau_{int}$}  \\
\hline\hline
$a$-$\eta^\prime$ & $\leq 150$     & \multicolumn{1}{|c|}{-}   \\
\hline
gluino-glue    & 620(60)    & 1100(200)                        \\
\hline
plaquette     & 378(37)    &  675(43)                          \\
\hline\hline
\end{tabular}
\end{center}
\end{table}
%%%%%%%%%%%%%%%%%%%%%%%%%%%%%%%%%%%%%%%%%%%%%%%%%%%%%%%%%%%%%%%%%%%%%%

%%%%%%%%%%%%%%%%%%%%%%%%%%%%%%%%%%%%%%%%%%%%%%%%%%%%%%%%%%%%%%%%%%%%%%%%
\section{Confinement potential}\label{sec3}

 The potential between static colour sources in gauge field theory is a
 physically very interesting quantity because it is characteristic for
 the dynamics of the gauge fields.
 If the sources are in the fundamental representation of the gauge group
 they can be called {\em static quarks}.

 For a model containing dynamical matter fields in the fundamental
 representation, as is the case for QCD with dynamical quarks, they will
 screen the static quarks.
 The potential then approaches a constant at large distances \cite{FS}.
 The string tension $\sigma$, which is the asymptotic slope of the
 potential for large distances, vanishes accordingly.
 This type of screening is of a more kinematical nature.

 On the other hand, if only matter fields in the adjoint representation
 of the gauge group are present, as in the case of supersymmetric
 N=1 Yang-Mills theory, there are different possibilities.
 Either the string tension does not vanish and static quarks are
 confined, or the static quarks are screened dynamically by the gauge
 fields.
 The latter situation is found in two-dimensional supersymmetric
 Yang-Mills theory \cite{2dSYM}.
 The screening mechanism is related to the chiral anomaly and appears to
 be specific to two dimensions.

 Four-dimensional SUSY Yang-Mills theory is believed to confine static
 quarks \cite{SHIFMAN}.
 Furthermore, the behaviour of the string tension as a function of the
 gluino mass can give indications on the question, whether QCD and
 Super-QCD are smoothly linked \cite{STRASSLER}.

 We have determined the static quark potential and the string tension
 for N=1 SUSY Yang-Mills theory from our Monte Carlo results.
 The starting point are expectation values of rectangular Wilson loops
 $\langle W(R,T) \rangle$.
 In order to improve the overlap with the relevant ground state we have
 applied APE-smearing \cite{APE} to the Wilson loops.
 The optimal smearing radius turns out to be near $R_s =3$.

\begin{figure}[htb]
\begin{center}
\epsfig{file=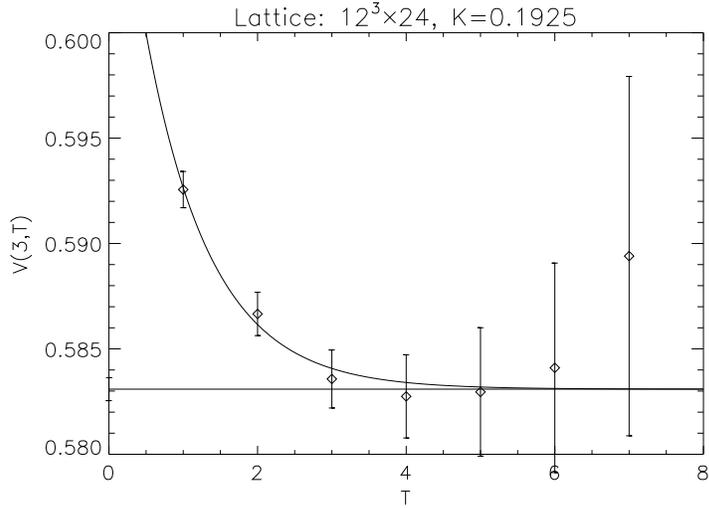,
        width=10cm}
\end{center}
\vspace{-3.0em}
\begin{center}
\parbox{15cm}{\caption{\label{Tfit_fig}\em
   Potential $V(R,T)$ for $R=3$ as a function of $T$ on a $12^3 \cdot
   24$ lattice. The line is an exponential fit to the large $T$
   behaviour, fitted over the range $1 \leq T \leq 6$.
}}
\end{center}
\end{figure}

 From the Wilson loops the potential can be found via
\be\label{eq3_1}
V(R) = \lim_{T \to \infty} V(R,T) \ ,
\ee
 where
\be\label{eq3_2}
V(R,T) = 
\log \langle W(R,T) \rangle - \log \langle W(R,T+1) \rangle \ .
\ee
 The large $T$ limit is approached exponentially \cite{BSS}.
 We have obtained the potential $V(R)$ through a fit of the form
\be\label{eq3_3}
V(R,T) = V(R) + c_1(R) e^{-c_2(R) T}.
\ee
 As an example we show $V(3,T)$ as a function of $T$ on a
 $12^3 \cdot 24$ lattice in figure \ref{Tfit_fig}.
 For $T > 6$ the errors grow significantly and we have chosen
 $1 \leq T \leq 6$ as the best fit interval on this lattice.
 On the $8^3 \cdot 16$ lattice fit intervals from $T=1$ to 4 or 5
 yield consistent results.

\begin{figure}[htb]
\begin{center}
\epsfig{file=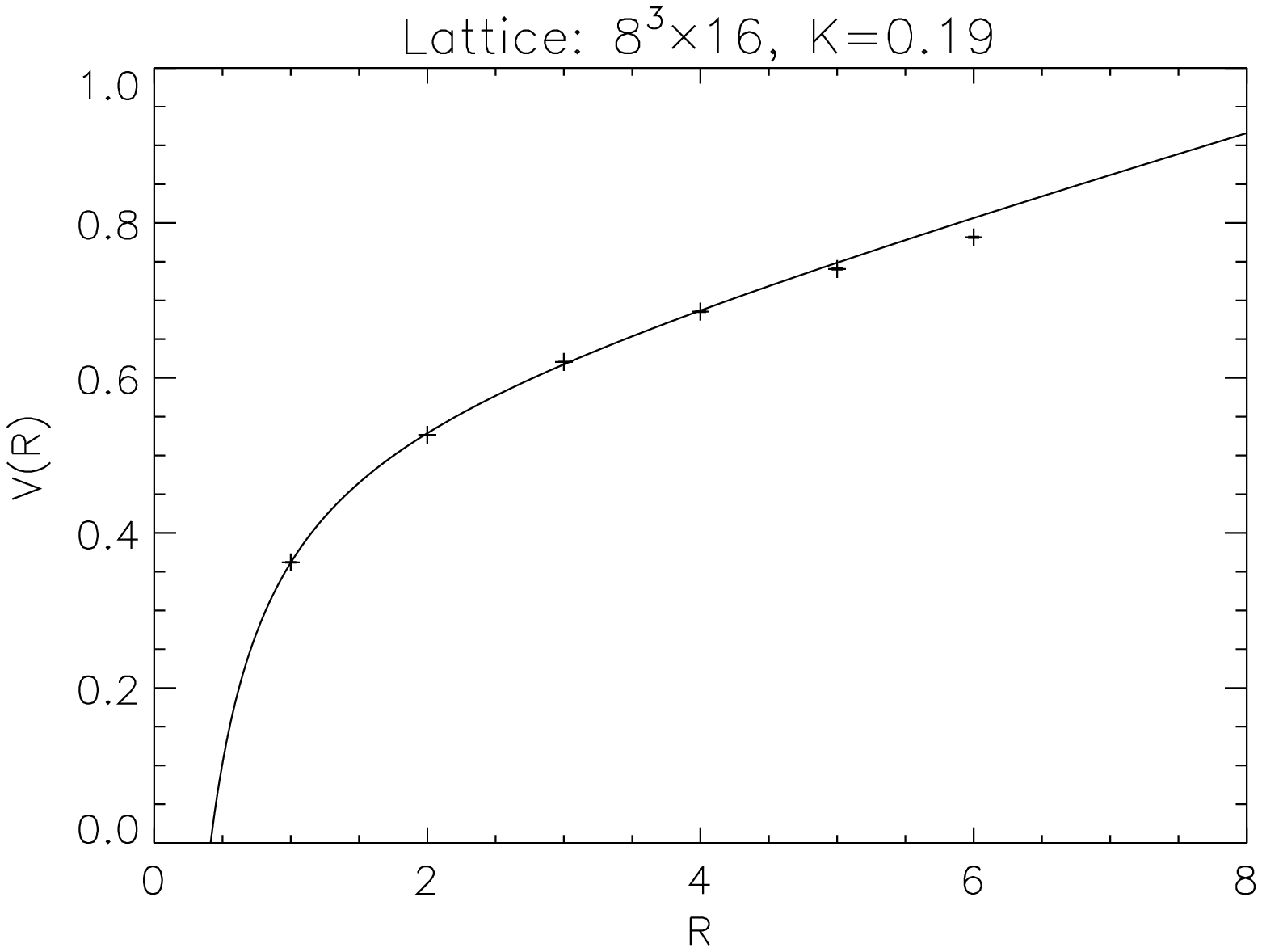,
        width=12cm}
\end{center}
\vspace{-3.0em}
\begin{center}
\parbox{15cm}{\caption{\label{VR8_fig}\em
   The static quark potential $V(R)$ on a $8^3 \cdot 16$ lattice at
   $K=0.19$. The line is a fit with a Coulomb plus a linear term,
   fitted over the range $1 \leq R \leq 4$.
}}
\end{center}
\end{figure}

\begin{figure}[htb]
\begin{center}
\epsfig{file=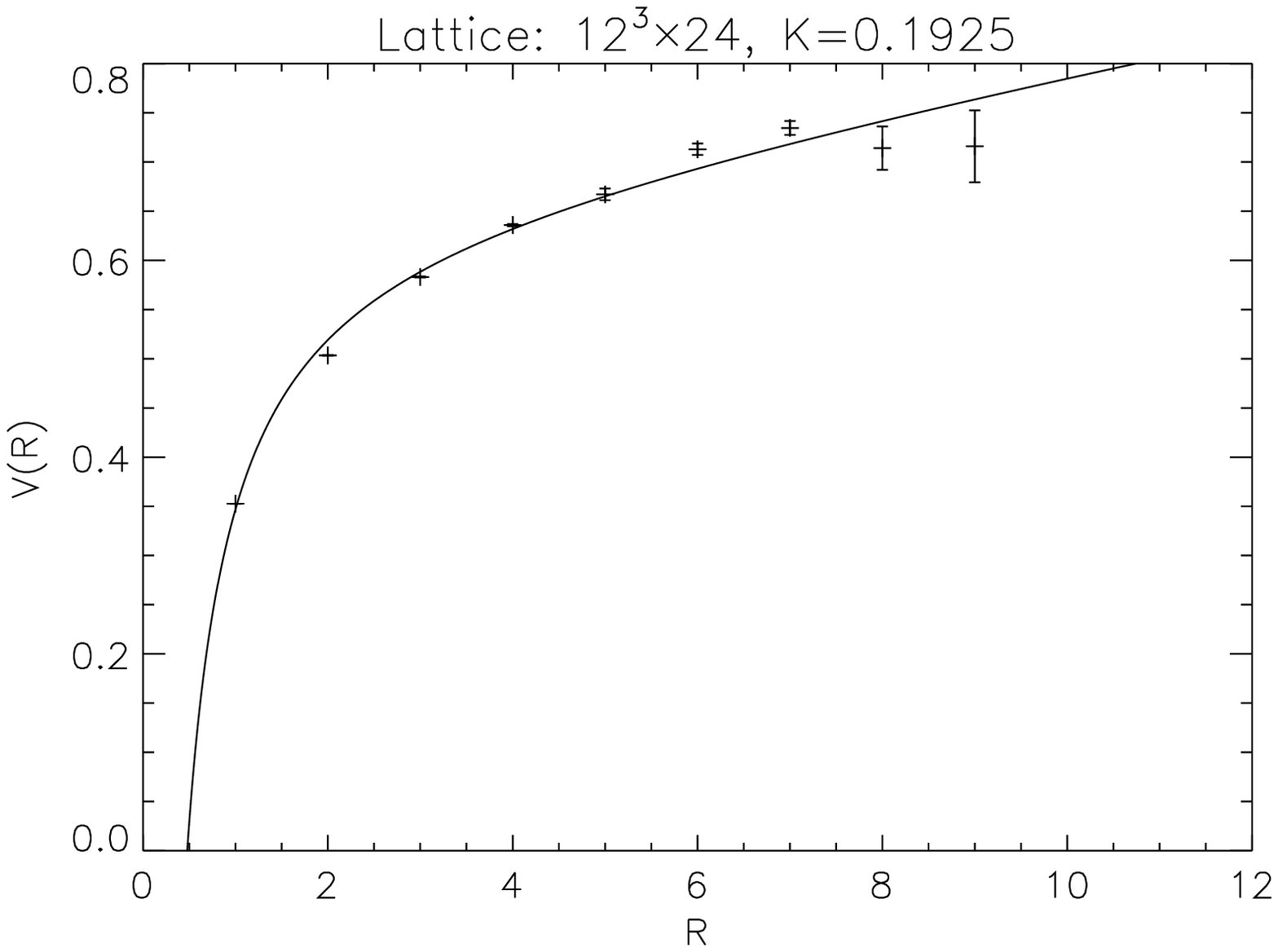,
        width=12cm}
\end{center}
\vspace{-3.0em}
\begin{center}
\parbox{15cm}{\caption{\label{VR_fig}\em
   The static quark potential $V(R)$ on a $12^3 \cdot 24$ lattice at
   $K=0.1925$. The line is a fit with a Coulomb plus a linear term,
   fitted over the range $1 \leq R \leq 6$.
}}
\end{center}
\end{figure}

 In this way the static potential $V(R)$ has been obtained for
 $1 \leq R \leq 6$ on the $8^3 \cdot 16$ lattice and for
 $1 \leq R \leq 9$ on the $12^3 \cdot 24$ lattice.
 For larger values of $R$ the errors become rather large and the results
 are not reliable anymore.
 Anyhow, for $R > L/2$ increasing finite size effects are to be
 expected.
 In figures~\ref{VR8_fig} and \ref{VR_fig} the potential is shown on
 the $L=8$ lattice at $K=0.19$ and $L=12$ lattice at $K=0.1925$,
 respectively.

 The string tension $\sigma$ is finally obtained by fitting the
 potential according to
\be\label{eq3_4}
V(R) = V_0 - \frac{\alpha}{R} + \sigma R.
\ee
 The value of $\sigma$ depends on the range of $R$ taken for the fit.
 In general it tends to decrease if the largest values of $R$ are
 included in the fit.
 However, this should not be interpreted as a signal for screening,
 since the potential is expected to bend down due to finite size
 effects.
 In table~\ref{Sigma_tab} the values for $\sqrt{\sigma}$ in
 lattice units are shown for different fit ranges.
%%%%%%%%%%%%%%%%%%%%%%%%%%%%%%%%%%%%%%%%%%%%%%%%%%%%%%%%%%%%%%%%%%%%%%%%
\begin{table}[htb]
\begin{center}
\parbox{15cm}{\caption{\label{Sigma_tab}\em
 Square root of the string tension $\sigma$ in lattice units and Coulomb
 strength $\alpha$ from fits to $V(R) = V_0-\frac{\alpha}{R}+\sigma R$
 over different ranges of $R$.
}}
\end{center}
\begin{center}
\begin{tabular}{| r | l | c | l | c |}
\hline
lattice & 
\multicolumn{1}{|c|}{$K$} & $R$ fit range &
\multicolumn{1}{|c|}{$a \sqrt{\sigma}$} & $\alpha$       \\
\hline\hline
$8^3 \cdot 16$ & 0.19 & 1 -- 4 & 0.22(1) & 0.23(2)       \\
$8^3 \cdot 16$ & 0.19 & 1 -- 5 & 0.21(1) & 0.25(1)       \\
$8^3 \cdot 16$ & 0.1925 & 1 -- 4 & 0.21(1) & 0.23(2)     \\
$8^3 \cdot 16$ & 0.1925 & 1 -- 5 & 0.19(1) & 0.25(2)     \\
$12^3 \cdot 24$ & 0.1925 & 1 -- 6 & 0.17(1) & 0.25(2)   \\
$12^3 \cdot 24$ & 0.1925 & 1 -- 7 & 0.16(1) & 0.26(2)    \\
$12^3 \cdot 24$ & 0.1925 & 1 -- 8 & 0.13(2) & 0.31(4)    \\
\hline
\end{tabular}
\end{center}
\end{table}
%%%%%%%%%%%%%%%%%%%%%%%%%%%%%%%%%%%%%%%%%%%%%%%%%%%%%%%%%%%%%%%%%%%%%%%%

 We consider the range $1 \leq R \leq L/2$ as reliable and quote as
 final results for the string tension
\begin{eqnarray}\label{eq3_5}
a \sqrt{\sigma} &=& 0.22(1) \hspace{5mm}  \mbox{for}\ K=0.1900, \ L=8,
\nonumber \\
a \sqrt{\sigma} &=& 0.21(1) \hspace{5mm}  \mbox{for}\ K=0.1925, \ L=8,
\nonumber \\
a \sqrt{\sigma} &=& 0.17(1) \hspace{5mm} \mbox{for}\ K=0.1925, \ L=12.
\end{eqnarray}
 The string tension in lattice units is decreasing when the critical
 line is approached, as it should be.
 This is mainly caused by the renormalization of the gauge coupling due
 to virtual gluino loop effects which are manifested by decreasing
 lattice spacing $a$.
 From a comparison of the $L=8$ and $L=12$ results one sees that finite
 size effects still appear to be sizable.
 This has to be expected because we have for the spatial lattice
 extension $L=12a$ the result $L\sqrt{\sigma} \simeq 2.1$.
 In QCD with $\sqrt{\sigma} \simeq 0.45\, GeV$ this would correspond
 to $L \simeq 1\, fm$.
 Although we are dealing with a different theory where finite size
 effects as a function of $L\sqrt{\sigma}$ are different, for a first
 orientation this estimate should be good enough.

 The coefficient $\alpha$ of the Coulomb term is close to the universal
 L\"uscher value of $\pi/12=0.26$ \cite{LUESYWEI}.

 For the ratio of the scalar glueball mass $m(0^+)$, to be discussed
 below, and the square root of the string tension we get
\begin{eqnarray}\label{eq3_6}
m(0^+) / \sqrt{\sigma} &=& 3.4(7) \hspace{3mm}
\mbox{for}\ K=0.1900, \ L=8,                       \nonumber \\
m(0^+) / \sqrt{\sigma} &=& 3.0(4) \hspace{3mm}
\mbox{for}\ K=0.1925, \ L=8,                       \nonumber \\
m(0^+) / \sqrt{\sigma} &=& 3.1(7) \hspace{3mm}
\mbox{for}\ K=0.1925, \ L=12.
\end{eqnarray}
 The uncertainties are not very small, but the numbers are consistent
 with a constant independent of $K$ in this range.
 They are of the same order of magnitude but somewhat smaller than in
 pure SU(2) gauge theory \cite{TEPER}, where at $\beta=2.5$ we have
 $m(0^+)/\sqrt{\sigma}=3.6$--$3.8$, depending on the lattice size.

%%%%%%%%%%%%%%%%%%%%%%%%%%%%%%%%%%%%%%%%%%%%%%%%%%%%%%%%%%%%%%%%%%%%%%%%
\section{Light bound state masses}\label{sec4}

 The non-vanishing string tension observed in the previous section is
 in accordance with the general expectation \cite{AKMRV,VENYAN} that
 the Yang-Mills theory with gluinos is confining.
 Therefore the asymptotic states are colour singlets, similarly to
 hadrons in QCD.
 The structure of the light hadron spectrum is closest to the
 (theoretical) case of QCD with a single flavour of quarks where the
 chiral symmetry is broken by the anomaly.

 Since both gluons and gluinos transform according to the adjoint
 (here triplet) representation of the colour group, one can construct
 colour singlet interpolating fields from any number of gluons and
 gluinos if their total number is at least two.
 Experience in QCD suggests that the lightest states can be well
 represented by interpolating fields build out of a small number of
 constituents.
 Simple examples are the {\em glueballs} known from pure Yang-Mills
 theory and {\em gluinoballs} corresponding to pseudoscalar mesons.
 We shall call the simplest pseudoscalar gluinoball made out of two
 gluinos the $a$-$\eta^\prime$ state.
 Here the label $a$ reminds us to the fact that the constituents are
 in the adjoint representation and $\eta^\prime$ stands for the
 corresponding $\eta^\prime$-meson in QCD.
 Mixed {\em gluino-glueball} states can be composed of any number of
 gluons and any number of gluinos, in the simplest case just one of
 both.

 In general, one has to keep in mind that the classification of states
 by some interpolating fields has only a limited validity, because
 this is a strongly interacting theory where many interpolating fields
 can have important projections on the same state.
 Taking just the simplest colour singlets can, however, give a good
 qualitative description.

 In the supersymmetric limit at zero gluino mass $m_{\tilde{g}}=0$ the
 hadronic states should occur in supermultiplets.
 This restricts the choice of simple interpolating field combinations
 and leads to low energy effective actions in terms of them 
 \cite{VENYAN,FAGASCH}.
 For non-zero gluino mass the supersymmetry is softly broken and
 the hadron masses are supposed to be analytic functions of
 $m_{\tilde{g}}$.
 The linear terms of a Taylor expansion in $m_{\tilde{g}}$ are often
 determined by the symmetries of the low energy effective actions
 \cite{LINEARMASS}.

 The effective action of Veneziano and Yankielowicz \cite{VENYAN}
 describes a chiral supermultiplet consisting of the $0^-$ gluinoball
 $a$-$\eta'$, the $0^+$ gluinoball $a$-$f_0$, and a spin $\half$
 gluino-glueball.
 There is, however, no a priori reason to assume that glueball states
 are heavier than the members of the supermultiplet above.
 Therefore Farrar, Gabadadze and Schwetz \cite{FAGASCH} proposed an
 effective action which includes an additional chiral supermultiplet.
 This multiplet consists of a $0^+$ glueball, a $0^-$ glueball and
 another gluino-glueball.
 The effective action allows mass mixing between the members of the two
 supermultiplets.
 The masses of the lightest bound states and the mixing among them can
 be investigated by Monte Carlo simulations.

%%%%%%%%%%%%%%%%%%%%%%%%%%%%%%%%%%%%%%%%%%%%%%%%%%%%%%%%%%%%%%%%%%%%%%%%
\subsection{Glueballs}\label{sec4.1}

 The glueball states as well as the methods to compute their masses
 in numerical Monte Carlo simulations are well known from pure gauge
 theory.
 (For a recent summary of results and references see \cite{TEPER}.)

 The lightest state is the $J^P=0^+$ glueball which can be generated
 by the symmetric combination of space-like plaquettes touching a
 lattice point.
 In order to optimize the signal and enhance the weight of the lightest
 state one is taking blocked \cite{TEPERBLOCK} or smeared \cite{APE}
 links instead of the original ones.
 In order to obtain the masses, for a first orientation, one can 
 use effective masses $m(t_1,t_2,T)$ assuming the dominance of a single
 state for time-slices $t_1,t_2$ on the periodic lattice with time
 extension $T$.
 One can search for time distance intervals where the effective
 masses are roughly constant and then try single mass fits in these
 intervals.
 In cases with high enough statistics and corresponding small
 statistical errors two-mass fits in larger intervals can also be
 stable and give information on the mass of the next excited state.

 Since no previous results on the glueball mass spectrum with dynamical
 gluinos are available in the literature, we started our search
 for dynamical gluino effects on small lattices as $4^3 \cdot 8$ at
 hopping parameter values $K \geq 0.16$.
 We observed some effects for $K \geq 0.18$ where we started
 runs on larger lattices, up to $12^3 \cdot 24$.
 As already seen in the previous section, the lattice spacing $a$ is
 decreasing with increasing $K$ (i.e.~decreasing gluino mass).
 This means that effectively we are closer to the continuum limit at
 larger $K$, resulting in smaller glueball (and other) masses in
 lattice units.
 This effect is strongest at zero gluino mass where a first order
 phase transition is expected due to the discrete chiral symmetry
 breaking.
 First numerical evidence for this phase transition has been reported
 by our collaboration at $K=K_0=0.1955(5)$ \cite{PHASETRANS}.

 With our spectrum calculations we stayed below this value and stopped
 at $K=0.1925$ where the $12^3 \cdot 24$ lattice is already not very
 large.
 The obtained masses for the $0^+$ glueball in lattice units are
\begin{eqnarray}\label{eq4.1_1}
a m(0^+) &=& 0.95(10) \hspace{3mm} \mbox{for}\ K=0.1800, \ L=6,
\nonumber \\
a m(0^+) &=& 0.85(6) \hspace{5mm}  \mbox{for}\ K=0.1850, \ L=6,
\nonumber \\
a m(0^+) &=& 0.75(6) \hspace{5mm}  \mbox{for}\ K=0.1900, \ L=8,
\nonumber \\
a m(0^+) &=& 0.63(5) \hspace{5mm}  \mbox{for}\ K=0.1925, \ L=8,
\nonumber \\
a m(0^+) &=& 0.53(10) \hspace{3mm} \mbox{for}\ K=0.1925, \ L=12.
\end{eqnarray}

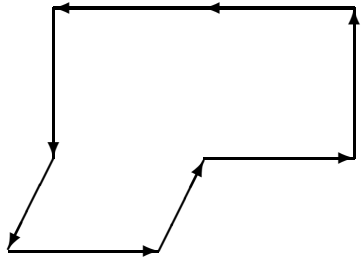
\begin{figure}[htb]
\begin{center}
\unitlength1cm
\begin{picture}(4,3)
\thicklines
\put(0,3){\vector(0,-1){2}}
\put(2,3){\vector(-1,0){2}}
\put(4,3){\vector(-1,0){2}}
\put(4,1){\vector(0,1){2}}
\put(2,1){\vector(1,0){2}}
\put(0,1){\vector(-1,-2){0.6}}
\put(-0.6,-0.235){\vector(1,0){2}}
\put(1.4,-0.235){\vector(1,2){0.6}}
\end{picture}
\end{center}
\vspace{-0.5em}
\begin{center}
\parbox{15cm}{\caption{\label{Loop_fig}\em
   Closed loop, which has been used to build the pseudoscalar glueball
   operator.
}}
\end{center}
\end{figure}

 In addition to the $J^P=0^+$ glueball we have studied the
 pseudoscalar $0^-$ glueball.
 In order to create a pseudoscalar glueball from the vacuum with an
 operator built from closed loops on the lattice, one needs loops which
 cannot be rotated into their mirror images.
 For gauge group SU(2) the traces of loop variables are real and
 do not distinguish the two orientations of loops.
 The smallest loops with the desired property are made of eight links.
 One possibility would be to take the simplest lattice version of
 Tr($\epsilon_{\mu \nu \rho\sigma} F^{\mu \nu} F^{\rho \sigma}$).
 However, it contains two orthogonal plaquettes and cannot be put into
 a single time-slice.
 Therefore we have chosen to take the loop $\mathcal{C}$ shown in
 figure~\ref{Loop_fig} \cite{BB}.

 The time-slice operator for the pseudoscalar glueball is then given by
\be\label{eq4.1_2}
S(t) = \sum_R [ {\rm Tr}\,U(\mathcal{C}) - {\rm Tr}\,U(P\mathcal{C})]
\ ,
\ee
 where the sum is over all rotations $R$ in the cubic lattice group and
 $P\mathcal{C}$ is the mirror image of $\mathcal{C}$.
 As usual, APE-smearing has been applied to the links appearing in the
 loop.

 The pseudoscalar glueball mass has been calculated from the time-slice
 correlation functions as an effective mass from distances 1 and 2 with
 optimized smearing radius.
 On the $6^3 \cdot 12$ lattice a good smearing radius is obtained for
 $R_s = 4$ or 5, and the numbers are very stable.
 On the $8^3 \cdot 16$ lattice a clear plateau in the number of
 smearing steps could not be seen.
 Nevertheless, for a smearing radius between 5 and 8 we obtain rather
 stable results.
 The masses in lattice units are
\begin{eqnarray}\label{eq4.1_3}
a m(0^-) &=& 1.5(3) \hspace{7mm} \mbox{for}\ K=0.1850, \ L=6,
\nonumber \\
a m(0^-) &=& 1.45(10) \hspace{3mm} \mbox{for}\ K=0.1900, \ L=6,
\nonumber \\
a m(0^-) &=& 1.3(1) \hspace{7mm} \mbox{for}\ K=0.1925, \ L=6,
\nonumber \\
a m(0^-) &=& 1.1(1) \hspace{7mm} \mbox{for}\ K=0.1925, \ L=8.
\end{eqnarray}
 The pseudoscalar glueball appears to be roughly twice as heavy as the
 scalar one.
 This is similar to pure SU(2) gauge theory, where
 $m(0^-)/m(0^+) = 1.8(2)$ \cite{TEPER}.

%%%%%%%%%%%%%%%%%%%%%%%%%%%%%%%%%%%%%%%%%%%%%%%%%%%%%%%%%%%%%%%%%%%%%%%%
\subsection{Gluino-glueballs}\label{sec4.2}

 One can construct colour singlet states from the gluinos and the field
 strength tensor in the adjoint representation.
 One of these states is a spin $\half$ Majorana fermion which occurs in
 the construction of the Veneziano-Yankielowicz effective
 action~\cite{VENYAN}.
 In order to find  the lowest mass in this channel we consider the
 correlator consisting of plaquettes connected by a quark propagator
 line:
\be\label{eq4.2_1}
\Gamma_{\tilde{g} g}(x,y) =
{\rm Tr_{spinor}}\,( \chi^r_x Q^{-1}_{xr,ys} \chi^s_y)
\ee
 where
\be\label{eq4.2_2}
\chi_x^r = \frac{1}{2i}{\rm Tr_{colour}}\,( \tau_r \bar{U}_x)
\ee
 and the plaquette variable is defined as
\be\label{eq4.2_3}
\bar{U}_x=U_{pl}(x,12)+U_{pl}(x,13)+U_{pl}(x,23) \ .
\ee
 For antiperiodic boundary conditions for the gluino in the time
 direction the correlator is antiperiodic.
 By inserting $\gamma_{4}$ into the correlation function (\ref{eq4.2_1})
 it becomes periodic also with antiperiodic boundary conditions.
 The resulting projection on the ground state have in both cases 
 either been compatible with one another, or the propagator modified
 with $\gamma_{4}$ has shown more mixing with larger masses. 
 Therefore in extracting the masses we considered only the above
 propagator without $\gamma_{4}$.

 For the gluino-glueball, in order to obtain a satisfactory signal,
 APE-smearing \cite{APE} has been implemented for the links and
 Jacobi-smearing \cite{UKQCD} for the gluino field.
 Tests have shown \cite{SPANDEREN} that Teper-blocking for the links
 was in this case not as well suited.
 Table~\ref{gluinoglu_smr_tab} shows the smearing parameters used for
 the gluino-glueball on different lattices at different hopping
 parameters.
 They have been optimized by measuring the masses on a small sample of
 data and  tuning the parameters accordingly to obtain the lowest mass
 values.
%%%%%%%%%%%%%%%%%%%%%%%%%%%%%%%%%%%%%%%%%%%%%%%%%%%%%%%%%%%%%%%%%%%%%%%%
\begin{table}[ht]
\begin{center}
\parbox{15cm}{\caption{\label{gluinoglu_smr_tab}\em
 Smearing parameters for Jacobi and APE-smearing used for measuring the 
 gluino-glueball.
}}
\end{center}
\begin{center}
\begin{tabular}{| c | l | c | l | c | l |}
\hline
 {\rm lattice} & \multicolumn{1}{|c|}{$K$} & $N_{Jacobi}$  &
\multicolumn{1}{|c|}{$K_{Jacobi}$}  &  $N_{APE}$  &
\multicolumn{1}{|c|}{$\epsilon_{APE}$}                        \\
\hline\hline
 $8^3 \cdot 16 $ &  0.19    &  20  &  0.22  &  8  & 0.35      \\ 
\hline
 $8^3 \cdot 16 $ &  0.1925  &  23  &  0.185 & 12  & 0.34      \\ 
\hline
$12^3 \cdot 24$  &  0.1925  &  19  &  0.20  &  9  & 0.3       \\
\hline\hline
\end{tabular}
\end{center}
\end{table}
%%%%%%%%%%%%%%%%%%%%%%%%%%%%%%%%%%%%%%%%%%%%%%%%%%%%%%%%%%%%%%%%%%%%%%%%

 The masses for the gluino-glueball were determined first by considering 
 effective masses $m(t_1,t_2,T)$ assuming the dominance of a single
 state for time-slices $t_1,t_2$ on the periodic lattice with time
 extension $T$.
 From this time distance intervals were determined where the effective
 masses were roughly constant and single mass fits in these intervals
 were performed.
 The results are shown in table~\ref{gluinoglu_res_tab}.
%%%%%%%%%%%%%%%%%%%%%%%%%%%%%%%%%%%%%%%%%%%%%%%%%%%%%%%%%%%%%%%%%%%%%%%%
\begin{table}[ht]
\begin{center}
\parbox{15cm}{\caption{\label{gluinoglu_res_tab}\em
 Lowest masses for the gluino-glueball at different hopping parameters
 and lattices.
 The value of the gauge coupling has been $\beta=2.3$ throughout.
}}
\end{center}
\begin{center}
\begin{tabular}{| c | l | l | l | l |}
\hline
\multicolumn{1}{|l|}{gluino-glueball}  &
\multicolumn{1}{|l|}{$K=0.18$ }        &
\multicolumn{1}{|l|}{$K=0.185$ }       &
\multicolumn{1}{|l|}{$K=0.19$ }        &
\multicolumn{1}{|l|}{$K=0.1925$ }                                     \\
\hline\hline
 $6^3 \cdot 12 $ & 1.93(5) & 1.39(8) & 1.05(20) &
 \multicolumn{1}{|c|}{-}                                              \\
\hline
 $8^3 \cdot 16$ & \multicolumn{1}{|c|}{-} & \multicolumn{1}{|c|}{-} &
 0.87(13) & 0.82(18)                                                  \\
\hline
$12^3 \cdot 24$ & \multicolumn{1}{|c|}{-} & \multicolumn{1}{|c|}{-} &
 \multicolumn{1}{|c|}{-}  &  0.93(8)                                  \\
\hline\hline
\end{tabular}
\end{center}
\end{table}
%%%%%%%%%%%%%%%%%%%%%%%%%%%%%%%%%%%%%%%%%%%%%%%%%%%%%%%%%%%%%%%%%%%%%%%%

%%%%%%%%%%%%%%%%%%%%%%%%%%%%%%%%%%%%%%%%%%%%%%%%%%%%%%%%%%%%%%%%%%%%%%%%
\subsection{Gluinoballs}\label{sec4.3}

 Besides the gluino-glueball in this work we consider also gluinoballs
 defined by a colourless combination of two gluino fields.
 The $a$-$\eta^\prime$ has spin-parity $0^-$ and the $a$-$f_0$
 spin-parity $0^+$.
 In the simulations for $a$-$\eta^\prime$ and $a$-$f_0$, respectively,
 the wave functions $\bar{\Psi}\gamma_5\Psi$ and $\bar{\Psi}\Psi$ were
 used.
 These gluinoballs are contained in the Veneziano-Yankielowicz 
 super-multiplet \cite{VENYAN}.
 For the correlation function a straightforward calculation as in
 \cite{TWO-STEP} with $\Gamma \in \{1,\gamma_5 \}$ yields
\be\label{eq4.3_1}
\Gamma_{\tilde {g} \tilde {g} } (x,y) =
\langle {\rm Tr_{sc}}\, \{ \Gamma Q^{-1}_{xx} \} 
{\rm Tr_{sc}}\, \{ \Gamma Q^{-1}_{yy} \} -
2\, {\rm Tr_{sc}}\, \{ \Gamma Q^{-1}_{xy} \Gamma Q^{-1}_{yx} \} \rangle
\ .
\ee
 Note the factor of two originating from the Majorana character of the
 gluinos.
 In analogy with a flavour singlet meson in QCD the propagator consists
 of a connected and a disconnected part: the left, respectively, the
 right term of (\ref{eq4.3_1}).

 The numerical evaluation of the time-slice of the connected part can be 
 reduced to the calculation of the propagator from a few initial points.
 The disconnected part is calculated using the volume source technique 
 \cite{KUFUMIOKUK}.
 For the determination of the gluinoball propagator no smearing has been
 used.

 In case of the $a$-$f_{0}$ particle the disconnected and the connected
 parts are of the same order of magnitude.
 The former has a much worse signal to noise ratio than the latter.
 This leads to a larger error on the $a$-$f_{0}$ as compared to the
 $a$-$\eta^\prime$ which is dominated by the connected part.   

 Our results for the $a$-$\eta^\prime$ and the $a$-$f_{0}$ masses for
 different  lattices and hopping parameters can be found in
 table~\ref{gluinoball_tab}.
%%%%%%%%%%%%%%%%%%%%%%%%%%%%%%%%%%%%%%%%%%%%%%%%%%%%%%%%%%%%%%%%%%%%%%%%
\begin{table}[ht]
\begin{center}
\parbox{15cm}{\caption{\label{gluinoball_tab}\em
 Lowest masses for the $a$-$\eta^\prime$ and the $a$-$f_{0}$ at
 different hopping parameters and lattices.
 The gauge coupling is given by $\beta=2.3$ throughout.
 In the last column with a star the next higher mass is shown, whenever
 it could be determined. 
}}
\end{center}
\begin{center}
\begin{tabular}{| l | l | l | l | l | l |}
\hline
\multicolumn{1}{|c|}{$a$-$\eta^\prime$}  &
\multicolumn{1}{|c|}{$K=0.18$}    &
\multicolumn{1}{|c|}{$K=0.185$}   &
\multicolumn{1}{|c|}{$K=0.19$}    &
\multicolumn{1}{|c|}{$K=0.1925$}  &
\multicolumn{1}{|c|}{$K=0.1925^*$ }                                   \\
\hline\hline
 $6^3 \cdot 12 $ & 1.155(11) & 0.941(8) & 0.594(14) &
 \multicolumn{1}{|c|}{-} &  \multicolumn{1}{|c|}{-}                   \\
\hline
 $8^3 \cdot 16 $ & \multicolumn{1}{|c|}{-} & \multicolumn{1}{|c|}{-} &
 0.725(20) & 0.551(17) & 1.282(26)                                    \\ 
\hline
$12^3 \cdot 24$ &  \multicolumn{1}{|c|}{-} & \multicolumn{1}{|c|}{-} &
 \multicolumn{1}{|c|}{-} & 0.48(5) & 1.09(5)                          \\
\hline\hline
\multicolumn{1}{|c|}{$a$-$f_{0}$}   &
\multicolumn{1}{|c|}{$K=0.18$}      &
\multicolumn{1}{|c|}{$K=0.185$}     &
\multicolumn{1}{|c|}{$K=0.19$ }     & 
\multicolumn{1}{|c|}{$K=0.1925$ }   &    \multicolumn{1}{|c|}{-}      \\
\hline\hline
 $6^3 \cdot 12 $ & 1.49(13) & 1.11(17) & \multicolumn{1}{|c|}{-} &
 \multicolumn{1}{|c|}{-} &                                            \\
\hline
 $8^3 \cdot 16 $ & \multicolumn{1}{|c|}{-} & \multicolumn{1}{|c|}{-} &
 1.20(22) & 0.81(17) &                                                \\
\hline
$12^3 \cdot 24$ & \multicolumn{1}{|c|}{-} & \multicolumn{1}{|c|}{-} &
 \multicolumn{1}{|c|}{-} &  1.00(13) &                                \\
\hline\hline
\end{tabular}
\end{center}
\end{table}
%%%%%%%%%%%%%%%%%%%%%%%%%%%%%%%%%%%%%%%%%%%%%%%%%%%%%%%%%%%%%%%%%%%%%%%%
 In case of the $a$-$\eta^\prime$ the data has been good enough to
 estimate also the next higher state.
 (These data can be found in the column denoted by a star.)
 The lowest masses have been obtained by using effective masses and
 fits as for the gluino-glueball.
 The fits were rather stable in case of the $a$-$\eta^\prime$ on the
 $12^3 \cdot 24 $ lattice.
 This allowed to extract two masses from the data.
 Errors were estimated by the jackknife method.

%%%%%%%%%%%%%%%%%%%%%%%%%%%%%%%%%%%%%%%%%%%%%%%%%%%%%%%%%%%%%%%%%%%%%%%%
\subsection{Glueball-gluinoball mixing}\label{sec4.4}

 In the low energy effective action of Farrar, Gabadadze and Schwetz
 \cite{FAGASCH} there is a possible non-zero mixing between the
 states in the two light supermultiplets.
 In particular there can be mixing of the $a$-$f_0$ gluinoball and the
 $0^+$ glueball.

 In order to study the mixing we have calculated the connected
 cross-correlation functions
\be\label{eq4.4_1}
\Gamma_{ij}(t) = \langle S_i(t_0) S_j(t_0 + t) \rangle_c
\ee
 where $i,j \in \{a,b\}$ and $S_a(t)$ is the plaquette operator
 creating a $0^+$ glueball from the vacuum, and $S_b(t)$ is the
 $\bar{\Psi}\Psi$ operator creating a $a$-$f_0$.
 If there is a non-zero mixing the hermitean correlation matrix
 $\Gamma_{ij}$ would not be diagonal.
 More generally one defines
\be\label{eq4.4_2}
\Lambda(t) = \left(
\begin{array}{cc}
\Gamma_{aa}(t) & \omega \Gamma_{ab}(t)\\
\omega \Gamma_{ba}(t) & \omega^2 \Gamma_{bb}(t)
\end{array}
\right),
\ee
 where $\omega$ is a real valued parameter.
 Diagonalizing $\Lambda(t)$ yields two eigenvalues, which are
 dominated by the lowest masses at large times $t$ \cite{LUEWOLF}:
\begin{eqnarray}\label{eq4.4_3}
\lambda_0(t)&=&f_0(\omega) e^{-m_0 t} \{ 1 + O( e^{-(m_1 - m_0)t}) \}\\
\lambda_1(t)&=&f_1(\omega) e^{-m_1 t} \{ 1 + O( e^{- \Delta m_1 t}) \},
\hspace{3mm} \Delta m_1 = \min(m_1 - m_0, m_2 - m_1).
\end{eqnarray}
 By tuning $\omega$ the statistical errors can be minimized.
 The masses $m_0$ and $m_1$ belong to the two lightest physical states
 in this channel.
 The mixing angle $\theta(t)$ is defined to be the angle between the
 eigenvector $v_0(t)$ corresponding to $\lambda_0$ and the vector
 $(1,0)$.
 For large $t$ one should observe a plateau where the mixing angle is
 constant and independent of $\omega$.

\begin{figure}[htb]
\begin{center}
\epsfig{file=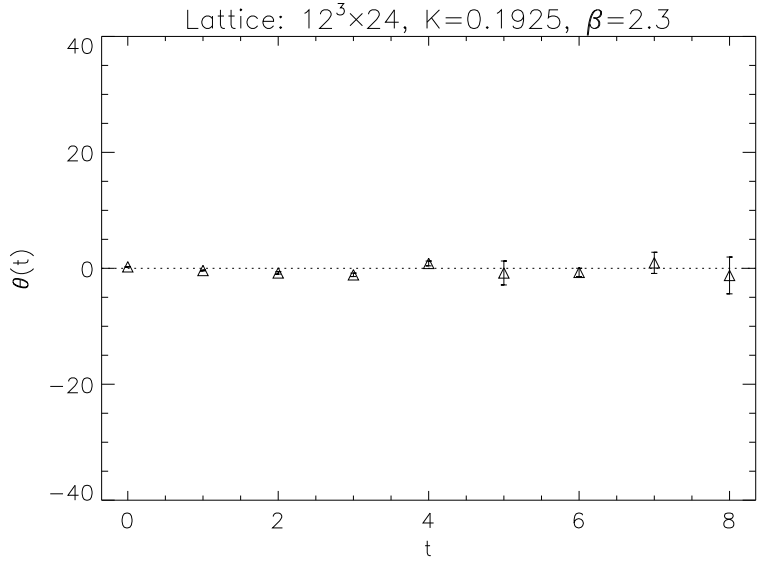,
        width=11cm}
\end{center}
\vspace{-3.5em}
\begin{center}
\parbox{15cm}{\caption{\label{mix_fig}\em
 The mixing angle $\theta(t)$ in the $0^+$ channel on a $12^3 \cdot
 24$ lattice at $K=0.1925$.
}}
\end{center}
\end{figure}

 We have determined the mixing angle in the $0^+$ channel from our Monte
 Carlo data.
 If $\omega$ takes its optimal value
 $\omega_0=\sqrt{\Gamma_{aa}/\Gamma_{bb}}$ \cite{UKQCD}, the errors are
 smallest.
 Figure~\ref{mix_fig} shows the mixing angle $\theta(t)$ for this
 choice of $\omega$.
 On the $8^3 \cdot 16$ lattice for $K=0.19$ and $K=0.1925$ as well as on
 the $12^3 \cdot 24$ lattice for $K=0.1925$ the result is consistent
 with zero within rather small errors.
 So there is no mixing between the glueball and the $a$-$f_0$ state.
 It might be possible that mixing only becomes visible in the close
 vicinity of the critical line corresponding to zero gluino mass, where
 supersymmetry is nearly restored.
 On the other hand, the effective action of \cite{FAGASCH} does not
 necessarily require a non-zero mixing to be present.

%%%%%%%%%%%%%%%%%%%%%%%%%%%%%%%%%%%%%%%%%%%%%%%%%%%%%%%%%%%%%%%%%%%%%%%%
\section{Summary and outlook}\label{sec5}

 The numerical Monte Carlo simulations presented in \cite{PHASETRANS}
 and this paper are the first calculations of this kind in a Yang-Mills
 theory with light gluinos.
 Therefore an essential part of our work had to be invested in
 algorithmic studies and parameter tuning.

 The two-step multi-bosonic algorithm, after appropriate tuning, turned
 out to be reliable and showed a satisfactory performance in the present
 case which is described by a flavour number $N_f=\half$ of fermions in
 the adjoint representation.
 We showed that the sign of the Pfaffian appearing in a path integral
 formulation of gluinos can be taken into account, but does not
 practically influence the results in the investigated range of
 parameters.
 Since the two-step multi-bosonic algorithm can also be applied for
 any number of fermion flavours in the fundamental representation, an
 interesting physical application would be, for instance, QCD with
 three light flavours of quarks.
 On the basis of our positive experience with the algorithm we expect
 that it would also work well in that case.

 Concerning parameter tuning in the lattice action, the problem is to
 find a region of bare parameter space where the gluino is light and
 where the lattice spacing is appropriate for feasible numerical
 simulations.
 Our strategy was to start at the lower end of the approximate scaling
 region in pure SU(2) lattice gauge theory at $\beta=2.3$ and to
 increase the hopping parameter $K$ as long as substantial effects of
 virtual dynamical gluino loops appear.
 It is expected that these effects decrease the lattice spacing due
 to the difference of the Callan-Symanzik $\beta$-functions with and
 without light gluinos.
 The observed effect is mainly an overall renormalization of $a$.
 The change of dimensionless ratios of masses and string tension are
 only moderate up to $K \leq 0.1925$, where most of our simulations
 were performed.

 Increasing $K$ further is getting more difficult from the algorithmic
 point of view because the smallest eigenvalues of the fermion matrix
 are becoming really small.
 In spite of this our algorithm still performed reasonably well.
 A search in the range up to $K \leq 0.20$ revealed first evidence
 for a first order phase transition expected to occur at zero gluino
 mass \cite{PHASETRANS}.
 Our present estimate for the location of this phase transition, at
 $\beta=2.3$, is $K_0=0.1955(5)$.
 This gives for the bare gluino mass in lattice units
\be\label{eq5_1}
am_0 \equiv \half \left[ \frac{1}{K}-\frac{1}{K_0} \right]
\ee
 a value $am_0 \simeq 0.04$ at $K=0.1925$.
 With the value of the string tension in (\ref{eq3_5}) we get
 $m_0/\sqrt{\sigma} \simeq 0.2$.
 Using QCD-units and neglecting the mass renormalization factor
 $Z_m$ of the order of 1 this corresponds to a light gluino mass of
 about $100\, MeV$.
 Of course, this can only serve as an order of magnitude guide because
 SYM and QCD are after all two different theories.
 In order to connect $m_0$ to, say, $\Lambda_{\overline{MS}}$ one had to
 perform a calculation as in \cite{ALPHA} with massless gluinos.

\begin{figure}[htb]
\begin{center}
\epsfig{file=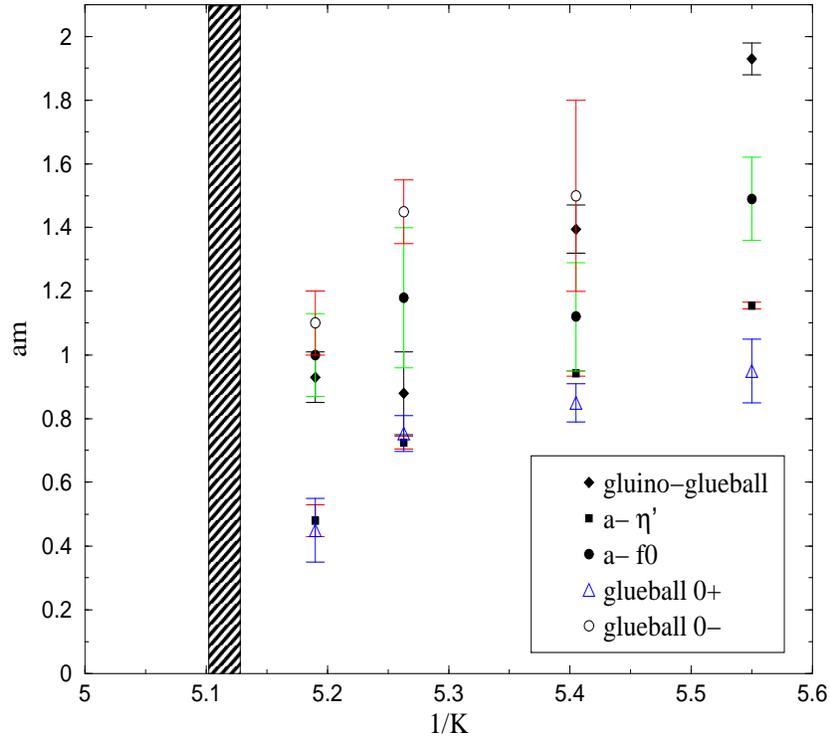,
        width=10cm,height=11cm,angle=270}
\end{center}
\vspace{-2.7em}
\begin{center}
\parbox{15cm}{\caption{\label{mass_fig}\em
 The lightest bound state masses in lattice units as function of the
 bare gluino mass parameter $1/K$.
 The shaded area at $K=0.1955(5)$ is where zero gluino mass and
 supersymmetry are expected.
}}
\end{center}
\end{figure}

%%%%%%%%%%%%%%%%%%%%%%%%%%%%%%%%%%%%%%%%%%%%%%%%%%%%%%%%%%%%%%%%%%%%%%
\begin{table}[ht]
\begin{center}
\parbox{15cm}{\caption{\label{mass_tab}\em
 Masses of the light bound states at $\beta=2.3$ in lattice units.
}}
\end{center}
\begin{center}
\begin{tabular}{| c | l | l | c | l | l | l |}
\hline
lattice: $L,T$  &
\multicolumn{1}{|c|}{$K$} &
\multicolumn{1}{|c|}{$am^{0^+}_{gg}$}  & $am^{0^-}_{gg} $ &
\multicolumn{1}{|c|}{$am^{0^+}_{\tilde{g}\tilde{g}}$} & 
\multicolumn{1}{|c|}{$am^{0^-}_{\tilde{g}\tilde{g}}$} & 
\multicolumn{1}{|c|}{$am_{\tilde{g}g}$}                             \\
\hline
6,12 &0.18  &0.95(10) & - & 1.49(13) & 1.155(11) & 1.93(5)          \\
\hline
6,12 &0.185 &0.85(6) & 1.5(3) & 1.11(17) & 0.941(8) & 1.39(8)       \\
\hline
8,16 &0.19 &0.75(6) & - & 1.20(22) & 0.725(20) & 0.87(13)           \\
\hline
8,16 &0.1925 &0.63(5) &1.1(1)&0.81(17) & 0.551(17) & 0.82(18)       \\
\hline
12,24 &0.1925 & 0.53(10) & -  &1.00(13) & 0.48(5) &0.93(8)          \\
\hline
\end{tabular}
\end{center}
\end{table}
%%%%%%%%%%%%%%%%%%%%%%%%%%%%%%%%%%%%%%%%%%%%%%%%%%%%%%%%%%%%%%%%%%%%%%
 Having an algorithm and knowing the interesting range of parameters
 in the lattice action one can start to perform numerical simulations
 for determining the spectrum of states and other physically
 interesting features.
 The properties of the lightest states are obviously quite interesting
 because the construction of low energy effective actions
 \cite{VENYAN,FAGASCH} is based on the assumptions about the relevant
 composite field variables.
 An important constraint on the spectrum is that in the limit of zero
 gluino mass, where supersymmetry is expected, the particle states
 should occur in supermultiplets with degenerate mass.
 A collection of our present results on the lightest states is
 displayed in table~\ref{mass_tab} and figure~\ref{mass_fig}.

 As one can see, for the lightest gluino masses (highest hopping
 parameters $K$) the bound state masses can be arranged into two groups.
 The lightest states are the $0^-$ gluinoball ($a$-$\eta^\prime$) and
 the $0^+$ glueball.
 At $K=0.1925$ these are in lattice units both near $am \simeq 0.5$.
 The other group of states is at $K=0.1925$ near $am \simeq 1.0$ and
 consists of the $0^+$ gluinoball, the $0^-$ glueball and the
 spin-$\half$ gluino-glueball.
 As shown by figure~\ref{mix_fig}, there is practically no mixing
 between the $0^+$-states in the two groups.
 The disturbing fact concerning supersymmetry is that there is
 apparently no spin-$\half$ state in the lower mass group.
 We saw this problem already at early stages of our project and hence
 paid specific attention to a lighter spin-$\half$ gluino-glueball
 state, but we did not find it.

 There are several possible explanations for this.
 Perhaps we are not yet close enough to the supersymmetric limit and
 therefore the spectrum does not yet look like a weakly broken
 supersymmetric spectrum.
 Another possibility is that we are missing the other spin-$\half$
 state because our choice of interpolating fields is not appropriate.
 One can, for instance, think about spin-$\half$ gluinoballs made out
 of three gluinos which appear at strong coupling~\cite{GAGOPE} and
 were not exploited in our simulations.
 Nevertheless, even if the spin-$\half$ state completing the lightest
 supermultiplet would be dominated by three gluinos, the emerging
 structure of the two light supermultiplets would be surprising.
 Finally one can think about possible finite volume effects and the
 effect of lattice artifacts breaking supersymmetry at finite lattice
 spacing.
 Without further numerical simulations we cannot exclude this last
 possibility but we believe that it is unlikely on basis of the
 experience in pure gauge theories.
 The product of the lattice spacing with the square-root of the string
 tension is at $K=0.1925$ given by $a\sqrt{\sigma} \simeq 0.17$.
 In pure SU(2) gauge theory \cite{TEPER} we have a similar value at
 $\beta \simeq 2.5-2.6$ which is within the region of reasonably good
 scaling.
 As discussed in section~\ref{sec3}, the spatial volume extension
 of our $12^3$ lattice at $K=0.1925$ is about $1\, fm$ in QCD units.
 This is almost certainly not large enough and therefore there are
 important finite volume effects to be expected, but the qualitative
 features of the bound state spectrum should already be visible in such
 volumes.

 We leave this puzzle for further investigations.
 The most important outcome of this first Super-Yang-Mills simulation
 with light gluinos is that the numerical Monte Carlo calculations are
 definitely possible with present-day techniques and can certainly
 contribute to the better understanding of the low energy
 non-perturbative dynamics of supersymmetric gauge theories.

%%%%%%%%%%%%%%%%%%%%%%%%%%%%%%%%%%%%%%%%%%%%%%%%%%%%%%%%%%%%%%%%%%%%%%%%
\vspace*{0.5em}
 {\bf Acknowledgement:}
 The numerical simulations presented here have been performed on the
 CRAY-T3E computers at John von Neumann Institute for Computing (NIC),
 J\"ulich.
 We thank NIC and the staff at ZAM for their kind support.

%%%%%%%%%%%%%%%%%%%%%%%%%%%%%%%%%%%%%%%%%%%%%%%%%%%%%%%%%%%%%%%%%%%%%%%%

%%%%%%%%%%%%%%%%%%%%%%%%%%%%%%%%%%%%%%%%%%%%%%%%%%%%%%%%%%%%%%%%%%%%%%%%
\newpage
\appendix\begin{center}  {\Large\bf Appendix}  \end{center}

\section{Least-squares optimized polynomials}\label{appenA}

 Least-squares optimization provides a general and flexible framework
 for obtaining the necessary optimized polynomials in multi-bosonic
 fermion algorithms.
 By exploiting different weight functions this framework is well suited
 to fulfill rather different requirements.

 In the first part of this appendix the basic formulae from
 \cite{POLYNOM} are collected.
 In the second part a simple example is considered:  in case of an
 appropriately chosen weight function the least-squares optimized
 polynomials for the approximation of the function $x^{-\alpha}$ are
 expressed in terms of Jacobi polynomials.

%%%%%%%%%%%%%%%%%%%%%%%%%%%%%%%%%%%%%%%%%%%%%%%%%%%%%%%%%%%%%%%%%%%%%%%%
\subsection{Definition and basic relations}\label{appenA.1}

 The general theory of least-squares optimized polynomial approximations
 can be inferred from the literature \cite{FOXPARKER,RIVLIN}.
 Here we introduce the basic formulae in the way it has been done in
 \cite{POLYNOM} for the specific needs of multi-bosonic fermion
 algorithms.
 We shall keep the notations there, apart from a few changes which
 allow for more generality.

 We want to approximate the real function $f(x)$ in the interval
 $x \in [\epsilon,\lambda]$ by a polynomial $P_n(x)$ of degree $n$.
 The aim is to minimize the deviation norm
\be \label{apA_1}
\delta_n \equiv \left\{ N_{\epsilon,\lambda}^{-1}\;
\int_\epsilon^\lambda dx\, w(x)^2 \left[ f(x) - P_n(x) \right]^2
\right\}^\half \ .
\ee
 Here $w(x)$ is an arbitrary real weight function and the overall
 normalization factor $N_{\epsilon,\lambda}$ can be chosen by
 convenience, for instance, as
\be \label{apA_2}
N_{\epsilon,\lambda} \equiv
\int_\epsilon^\lambda dx\, w(x)^2 f(x)^2 \ .
\ee
 A typical example of functions to be approximated is
 $f(x)=x^{-\alpha}/\bar{P}(x)$ with $\alpha > 0$ and some polynomial
 $\bar{P}(x)$.
 The interval is usually such that $0 \leq \epsilon < \lambda$.
 For optimizing the relative deviation one takes a weight function
 $w(x) = f(x)^{-1}$.

 It turns out useful to introduce orthogonal polynomials
 $\Phi_\mu(x)\; (\mu=0,1,2,\ldots)$ satisfying
\be \label{apA_3}
\int_\epsilon^\lambda dx\, w(x)^2 \Phi_\mu(x)\Phi_\nu(x)
= \delta_{\mu\nu} q_\nu \ .
\ee
 and expand the polynomial $P_n(x)$ in terms of them:
\be \label{apA_4}
P_n(x) = \sum_{\nu=0}^n d_{n\nu} \Phi_\nu(x) \ .
\ee
 Besides the normalization factor $q_\nu$ let us also introduce, for
 later purposes, the integrals $p_\nu$ and $s_\nu$ by
\be \label{apA_5}
q_\nu \equiv \int_\epsilon^\lambda dx\, w(x)^2 \Phi_\nu(x)^2 \ ,
\hspace{2em}
p_\nu \equiv \int_\epsilon^\lambda dx\, w(x)^2 \Phi_\nu(x)^2 x \ ,
\hspace{2em}
s_\nu \equiv \int_\epsilon^\lambda dx\, w(x)^2 x^\nu \ .
\ee

 It can be easily shown that the expansion coefficients $d_{n\nu}$
 minimizing $\delta_n$ are independent of $n$ and are given by
\be \label{apA_6}
d_{n\nu} \equiv d_\nu = \frac{b_\nu}{q_\nu} \ ,
\ee
 where
\be \label{apA_7}
b_\nu \equiv \int_\epsilon^\lambda dx\, w(x)^2 f(x) \Phi_\nu(x) \ .
\ee
 The minimal value of $\delta_n^2$ is
\be \label{apA_8}
\delta_n^2 = 1 - N_{\epsilon,\lambda}^{-1}
\sum_{\nu=0}^n d_\nu b_\nu \ .
\ee

 The above orthogonal polynomials satisfy three-term recurrence
 relations which are very useful for numerical evaluation.
 The first two of them with $\mu=0,1$ are given by
\be \label{apA_9}
\Phi_0(x) = 1 \ , \hspace{2em}
\Phi_1(x) = x - \frac{s_1}{s_0} \ .
\ee
 The higher order polynomials $\Phi_\mu(x)$ for $\mu=2,3,\ldots$ can be
 obtained from the recurrence relation
\be \label{apA_10}
\Phi_{\mu+1}(x) = (x+\beta_\mu)\Phi_\mu(x) +
\gamma_{\mu-1}\Phi_{\mu-1}(x) \ ,
\hspace{2em} (\mu=1,2,\ldots) \ ,
\ee
 where the recurrence coefficients are given by
\be \label{apA_11}
\beta_\mu = -\frac{p_\mu}{q_\mu} \ , \hspace{3em}
\gamma_{\mu-1} = -\frac{q_\mu}{q_{\mu-1}} \ .
\ee

 Using these relations on can set up a recursive scheme for the
 computation of the orthogonal polynomials in terms of the basic
 integrals $s_\nu$ defined in (\ref{apA_5}).
 Defining the polynomial coefficients $f_{\mu\nu}\; (0\leq\nu\leq\mu)$
 by
\be \label{apA_12}
\Phi_\mu(x) = \sum_{\nu=0}^\mu f_{\mu\nu}x^{\mu-\nu}
\ee
 the above recurrence relations imply the normalization convention
\be \label{apA_13}
f_{\mu 0} = 1 \ , \hspace{3em} (\mu=0,1,2,\ldots) \ ,
\ee
 and one can easily show that $q_\mu$ and $p_\mu$ satisfy
\be \label{apA_14}
q_\mu = \sum_{\nu=0}^\mu f_{\mu\nu} s_{2\mu-\nu} \ ,
\hspace{2em}
p_\mu = \sum_{\nu=0}^\mu f_{\mu\nu}
\left( s_{2\mu+1-\nu} +f_{\mu 1}s_{2\mu-\nu} \right) \ .
\ee
 The coefficients themselves can be calculated from $f_{11}=-s_1/s_0$
 and (\ref{apA_10}) which gives
\begin{eqnarray}\label{apA_15}
f_{\mu+1,1} & = & f_{\mu,1} + \beta_\mu \ ,             \nonumber \\
f_{\mu+1,2} & = & f_{\mu,2} + \beta_\mu f_{\mu,1} +
\gamma_{\mu-1} \ ,                                      \nonumber \\
f_{\mu+1,3} & = & f_{\mu,3} + \beta_\mu f_{\mu,2} +
\gamma_{\mu-1} f_{\mu-1,1} \ ,                          \nonumber \\
            & \ldots &                                  \nonumber \\
f_{\mu+1,\mu} & = & f_{\mu,\mu} + \beta_\mu f_{\mu,\mu-1} +
\gamma_{\mu-1} f_{\mu-1,\mu-2} \ ,                      \nonumber \\
f_{\mu+1,\mu+1} & = & \beta_\mu f_{\mu,\mu} +
\gamma_{\mu-1} f_{\mu-1,\mu-1} \ .
\end{eqnarray}
 The polynomial and recurrence coefficients are recursively determined
 by (\ref{apA_13})-(\ref{apA_15}).
 The expansion coefficients for the optimized polynomial $P_n(x)$
 can be obtained from (\ref{apA_6}) and
\be \label{apA_16}
b_\mu = \sum_{\nu=0}^\mu f_{\mu\nu}
\int_\epsilon^\lambda dx\, w(x)^2 f(x) x^{\mu-\nu} \ .
\ee
%

%%%%%%%%%%%%%%%%%%%%%%%%%%%%%%%%%%%%%%%%%%%%%%%%%%%%%%%%%%%%%%%%%%%%%%%%
\subsection{A simple example: Jacobi polynomials}\label{appenA.2}

 The approximation interval $[\epsilon,\lambda]$ can be transformed to
 some standard interval, say, $[-1,1]$ by the linear mapping
\be \label{apA_17}
\xi = \frac{2x-\lambda-\epsilon}{\lambda-\epsilon} \ , \hspace{2em}
x = \frac{\xi}{2}(\lambda-\epsilon) + \half(\lambda+\epsilon) \ .
\ee
 A weight factor $(1+\xi)^\rho(1-\xi)^\sigma$ with $\rho,\sigma > -1$
 corresponds in the original interval to the weight factor
\be \label{apA_18}
w^{(\rho,\sigma)}(x)^2 = (x-\epsilon)^\rho (\lambda-x)^\sigma \ .
\ee
 Taking, for instance, $\rho=2\alpha,\; \sigma=0$ this weight is similar
 to the one for relative deviation from the function $f(x)=x^{-\alpha}$,
 which would be just $x^{2\alpha}$.
 In fact, for $\epsilon=0$ these are exactly the same and for small
 $\epsilon$ the difference is negligible.
 The advantage of considering the weight factor in (\ref{apA_18}) is that
 the corresponding orthogonal polynomials are simply related to the
 Jacobi polynomials \cite{FREUD,RIGR}, namely
\be \label{apA_19}
\Phi^{(\rho,\sigma)}_\nu(x) = (\lambda-\epsilon)^\nu \nu!\,
\frac{\Gamma(\rho+\sigma+\nu+1)}{\Gamma(\rho+\sigma+2\nu+1)}
P^{(\sigma,\rho)}_\nu
\left(\frac{2x-\lambda-\epsilon}{\lambda-\epsilon}\right) \ .
\ee
 Our normalization convention (\ref{apA_13}) implies that
\be \label{apA_20}
q^{(\rho,\sigma)}_\nu = (\lambda-\epsilon)^{\rho+\sigma+2\nu+1}\nu!\,
\frac{\Gamma(\rho+\nu+1)\Gamma(\sigma+\nu+1)\Gamma(\rho+\sigma+\nu+1)}
{\Gamma(\rho+\sigma+2\nu+1)\Gamma(\rho+\sigma+2\nu+2)} \ .
\ee
 The coefficients of the orthogonal polynomials are now given by
\be \label{apA_21}
f^{(\rho,\sigma)}_{\mu\nu} = \sum_{\omega=0}^\nu
(-\epsilon)^{\nu-\omega}(\epsilon-\lambda)^\omega
\left(\begin{array}{c} \mu-\omega \\ \nu-\omega \end{array}\right)
\left(\begin{array}{c} \mu \\ \omega \end{array}\right)
\frac{\Gamma(\rho+\mu+1)\Gamma(\rho+\sigma+2\mu-\omega+1)}
{\Gamma(\rho+\mu-\omega+1)\Gamma(\rho+\sigma+2\mu+1)} \ .
\ee
 In particular, we have
\be \label{apA_22}
f^{(\rho,\sigma)}_{\mu0} = 1 \ , \hspace{2em}
f^{(\rho,\sigma)}_{11} = -\epsilon - (\lambda-\epsilon)
\frac{(\rho+1)}{(\rho+\sigma+2)} \ .
\ee
 The coefficients $\beta,\gamma$ in the recurrence relation (\ref{apA_10})
 can be derived from the known recurrence relations of the Jacobi
 polynomials:
$$
\beta^{(\rho,\sigma)}_\mu = -\half(\lambda+\epsilon) +
\frac{(\sigma^2-\rho^2)(\lambda-\epsilon)}
{2(\rho+\sigma+2\mu)(\rho+\sigma+2\mu+2)} \ ,
$$
\be \label{apA_23}
\gamma^{(\rho,\sigma)}_{\mu-1} = -(\lambda-\epsilon)^2\,
\frac{\mu(\rho+\mu)(\sigma+\mu)(\rho+\sigma+\mu)}
{(\rho+\sigma+2\mu-1)(\rho+\sigma+2\mu)^2(\rho+\sigma+2\mu+1)} \ .
\ee

 In order to obtain the expansion coefficients of the least-squares
 optimized polynomials one has to perform the integrals in (\ref{apA_16}).
 As an example, let us consider the function $f(x)=x^{-\alpha}$ when
 the necessary integrals can be expressed by hypergeometric functions:
$$
\int_\epsilon^\lambda dx\, (x-\epsilon)^\rho (\lambda-x)^\sigma
x^{\mu-\nu-\alpha} =
$$
\be \label{apA_24}
= (\lambda-\epsilon)^{\rho+\sigma+1} \lambda^{\mu-\nu-\alpha}
\frac{\Gamma(\rho+1)\Gamma(\sigma+1)}{\Gamma(\rho+\sigma+2)}
F\left( \alpha-\mu+\nu,\sigma+1;\rho+\sigma+2;
1-\frac{\epsilon}{\lambda} \right) \ .
\ee
 Let us now consider, for simplicity, only the case $\epsilon=0$, when
 we obtain
\be \label{apA_25}
b^{(\rho,\sigma)}_\mu = (-1)^\mu \lambda^{1+\rho+\sigma+\mu-\alpha}
\frac{\Gamma(\rho+\sigma+\mu+1)\Gamma(\alpha+\mu)\Gamma(\rho-\alpha+1)
\Gamma(\sigma+\mu+1)}{\Gamma(\rho+\sigma+2\mu+1)\Gamma(\alpha)
\Gamma(\rho+\sigma-\alpha+\mu+2)} \ .
\ee
 Combined with (\ref{apA_6}) and (\ref{apA_20}) this leads to
\be \label{apA_26}
d^{(\rho,\sigma)}_\mu = (-1)^\mu \lambda^{-\mu-\alpha}
\frac{\Gamma(\rho+\sigma+2\mu+2)\Gamma(\alpha+\mu)\Gamma(\rho-\alpha+1)}
{\mu!\,\Gamma(\rho+\mu+1)\Gamma(\alpha)\Gamma(\rho+\sigma-\alpha+\mu+2)}
 \ .
\ee

 These formulae can be used, for instance, for fractional inversion.
 For the parameters $\rho,\sigma$ the natural choice in this case is
 $\rho=2\alpha,\sigma=0$ which corresponds to the optimization of the
 relative deviation from the function $f(x)=x^{-\alpha}$.
 As we have seen in section \ref{appenA.1}, the optimized polynomials
 are the truncated expansions of $x^{-\alpha}$ in terms of the Jacobi
 polynomials $P^{(2\alpha,0)}$.
 The Gegenbauer polynomials proposed in \cite{BUNK} for fractional
 inversion correspond to a different choice, namely
 $\rho=\sigma=\alpha-\half$.
 This is because of the relation
\be \label{apA_27}
C^\alpha_n(x) = \frac{\Gamma(n+2\alpha)\Gamma(\alpha+\half)}
{\Gamma(2\alpha)\Gamma(n+\alpha+\half)}
P^{(\alpha-\half,\alpha-\half)}_n(x) \ .
\ee
 Note that for the simple case $\alpha=1$ we have here the Chebyshev
 polynomials of second kind: $C^1_n(x)=U_n(x)$.

\begin{figure}[htb]
\begin{flushleft}
\epsfig{file=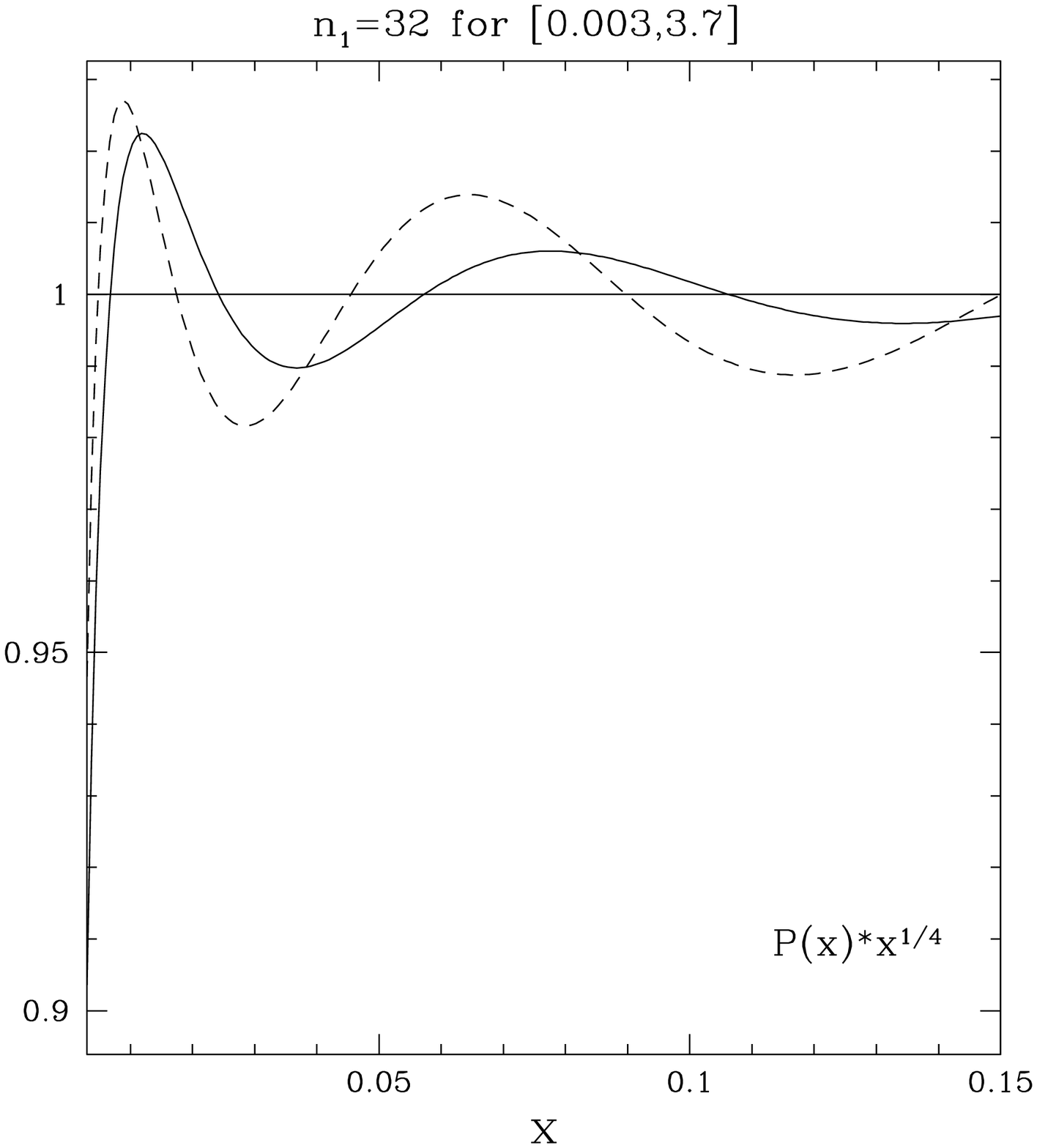,
        width=55mm,height=72mm}
\end{flushleft}
\vspace*{-82mm}
\begin{center}
\epsfig{file=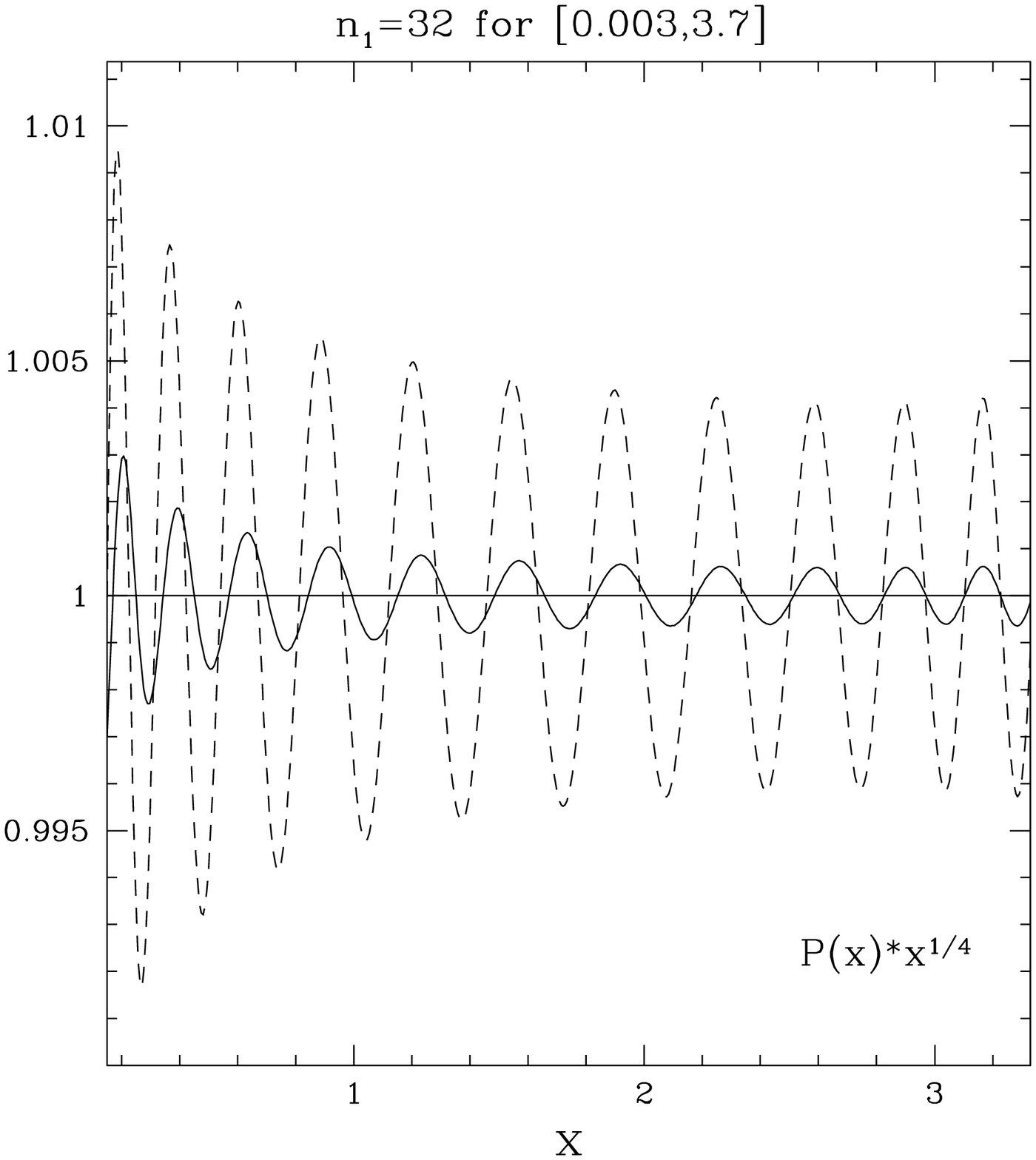,
        width=55mm,height=72mm}
\end{center}
\vspace*{-82mm}
\begin{flushright}
\epsfig{file=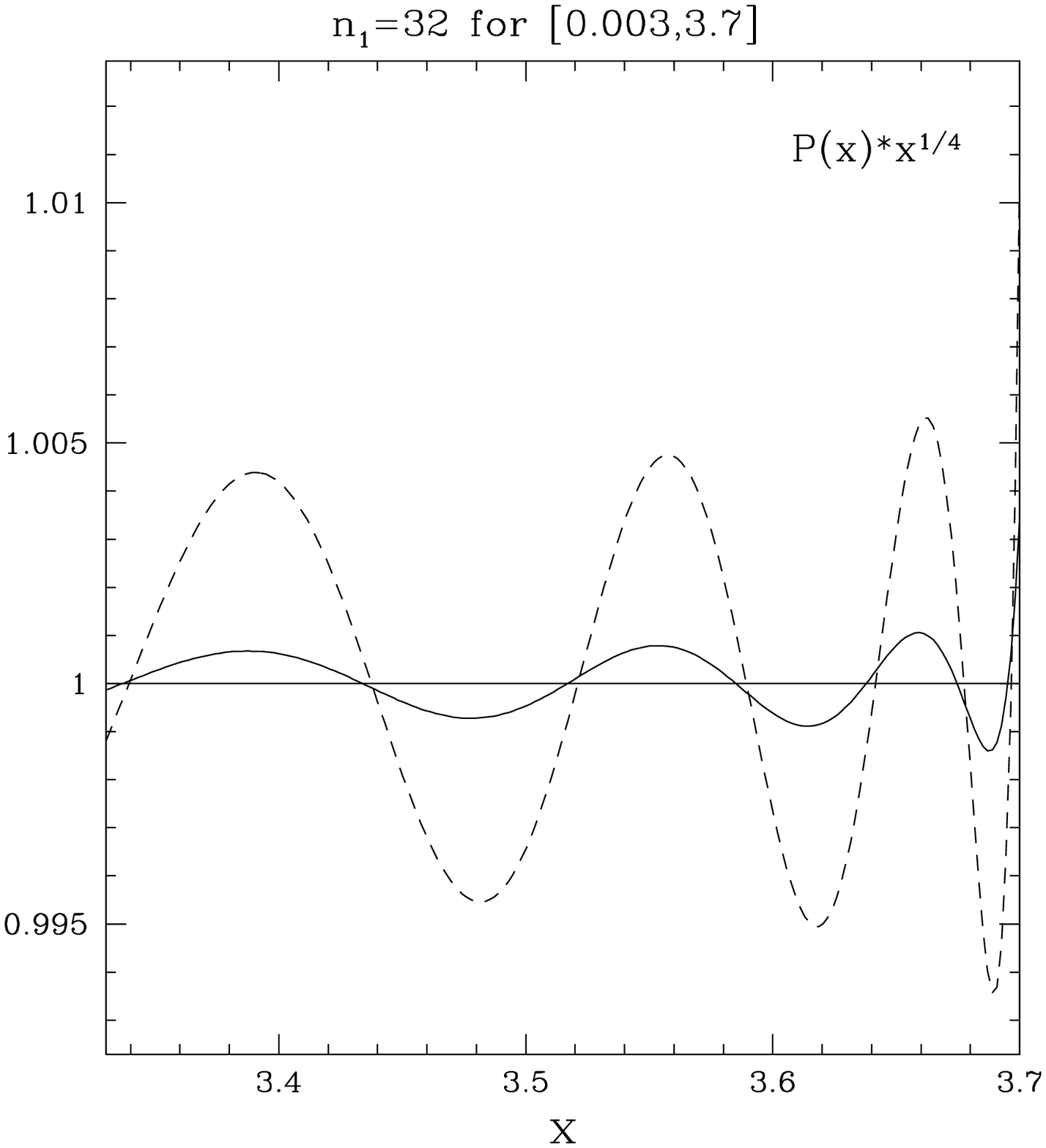,
        width=55mm,height=72mm}
\end{flushright}
\vspace*{-2.0em}
\begin{center}
\parbox{15cm}{\caption{\label{gegenb_fig}\em
 Comparing the polynomial approximations of $x^{-1/4}$ in the interval
 $[\epsilon,\lambda]=[0.003,3.7]$ at the order $n_1=32$.
 $P^{(1)}_{n_1}(x)\, x^{1/4}$ is shown for the least squares optimized
 polynomial minimizing the relative deviation (full line) and for the
 fractional inversion defined by the Gegenbauer polynomials with index
 $\alpha=\frac{1}{4}$ (dashed line).
 The interval is shown in three parts in order to display better the
 details.
}}
\end{center}
\end{figure}

 In our present application we have to consider $\alpha=\frac{1}{4}$.
 For the first polynomial $P^{(1)}_{n_1}$ we could, for instance, use
 the Gegenbauer polynomials $G^{\frac{1}{4}}$ corresponding to
 $P^{(-\frac{1}{4},-\frac{1}{4})}$.
 (For $P^{(2,3,4)}$ we need, of course, the polynomials introduced in
 \cite{TWO-STEP} which approximate more complicated functions.)
 A numerical comparison shows, however, that the least squares optimized
 polynomials minimizing the relative deviation in the interval
 $[\epsilon,\lambda]$ are better than the Gegenbauer polynomials (see
 fig.~\ref{gegenb_fig}): both approximations are similar at the lower
 end of the interval but otherwise the deviations of the former are by a
 factor of five smaller.

 The special case $\alpha=\half$ is interesting for the numerical
 evaluation of the zero mass lattice action proposed by Neuberger
 \cite{NEUBERGER}.
 In this case, in order to obtain the least-squares optimized relative
 deviation with weight function $w(x)=x$, the function $x^{-\half}$ has
 to be expanded in the Jacobi polynomials $P^{(1,0)}$.
 Note that this is different both from the Chebyshev and the Legendre
 expansions applied in \cite{HEJALU}.
 The former would correspond to take $P^{(-\half,-\half)}$, the latter
 to $P^{(0,0)}$.
 The corresponding weight functions would be $[x(\lambda-x)]^{-\half}$
 and $1$, respectively.
 As a consequence of the divergence of the weight factor at $x=0$,
 the Chebyshev expansion is not appropriate for an approximation in an
 interval with $\epsilon=0$.
 This can be immediately seen from the divergence of
 $d^{(-\half,-\half)}_\mu$ at $\alpha=\half$ in (\ref{apA_26}).

 The advantage of the Jacobi polynomials appearing in these examples is
 that they are analytically known.
 The more general least-squares optimized polynomials defined in the
 previous subsection can also be numerically expanded in terms of them.
 This is sometimes more comfortable than the entirely numerical
 approach.

%%%%%%%%%%%%%%%%%%%%%%%%%%%%%%%%%%%%%%%%%%%%%%%%%%%%%%%%%%%%%%%%%%%%%%%%
\section{Determining the smallest eigenvalues}\label{appenB}

 For finding the smallest eigenvalues and the corresponding eigenvectors
 of the squared hermitean fermion matrix
 $\tilde{Q}^2 = Q^\dagger Q$ we apply the algorithm of Kalkreuter and
 Simma \cite{KALKSIMM}.
 Some modifications and the optimization with detailed tests have been
 described in \cite{SPANDEREN}.
 Here we give a short summary for the readers convenience.

 The smallest eigenvalue of a general hermitean matrix $H$ can be found
 by minimizing the {\em Ritz functional}
\be \label{apB_1}
\mu_H(z) \equiv \frac{(z^* H z)}{(z^* z)} \ .
\ee
 Here the notation defined in (\ref{eq2.3_2}) is used, with $z^*$
 denoting the complex conjugate vector of $z$ and
 $(x y) \equiv (x I y)=(y x)$, where $I$ is the unit matrix.
 The gradient of the Ritz functional is obviously
\be \label{apB_2}
g_H(z) = \frac{Hz(z^* z) - z(z^* H z)}{(z^* z)^2} \ .
\ee
 For the {\em conjugate gradient procedure} we can choose a starting
 search direction $p_1 = -g_H(z)$ and the iteration is defined by a new
 approximation to the eigenvector
\be \label{apB_3}
z_{i+1} = z_i + \alpha_i p_i \ ,  \hspace{3em} i=1,2,\ldots\; \ .
\ee
 The factor $\alpha_i$ is chosen at the minimum of $\mu_H(z)$ in the
 search direction $p_i$.
 One can show that
\be \label{apB_4}
\alpha_i = \frac{2\, (p_i^* H z_i)}{(z_i^* H z_i) - (p_i^* H p_i)
- \sqrt{[(z_i^* H z_i) - (p_i^* H p_i)]^2
+ 4\, (p_i^* H z_i)(z_i^* H p_i)]}} \ .
\ee
 Let us note that taking the positive sign in front of the square root
 gives the maximum, instead of the minimum.
 The other sign can be used for finding the maximal eigenvalue instead
 of the minimal one.
 In the iteration relation (\ref{apB_3}) the conjugate search direction
 $p_{i+1}$ can be chosen according to \cite{KALKSIMM}
\be \label{apB_5}
p_{i+1} = g_H(z_{i+1}) + \beta_i \left[
p_i - z_{i+1}\frac{(z^*_{i+1} p_i)}{(z^*_{i+1} z_{i+1})}
\right] \ .
\ee
 For the factor $\beta_i$ one can take, according to the
 {\em Fletcher-Reeves} prescription, with $g_i \equiv g_H(z_i)$
\be \label{apB_6}
\beta_i = \frac{(g^*_{i+1} g_{i+1})}{(g^*_i g_i)}
\ee
 or alternatively, according to the {\em Polak-Ribiere} prescription,
\be \label{apB_7}
\beta_i = \frac{(g^*_{i+1} g_{i+1})-(g^*_i g_{i+1})}{(g^*_i g_i)} \ .
\ee
 It turned out that in case of our fermion matrices the Polak-Ribiere
 version is 25\% to 40\% more efficient than the Fletcher-Reeves
 version proposed in \cite{KALKSIMM}.
 In naive implementations of this iterative procedure numerical problems
 may occur due to the increasing length of the vector $z_i$.
 Since the Ritz functional is scale invariant, this problem can be
 avoided by rescaling, typically every 25 steps, as
\be \label{apB_8}
z_i \to \frac{z_i}{\sqrt{(z_i^* z_i)}} \ ,  \hspace{2em}
p_i \to p_i\, \sqrt{(z_i^* z_i)} \ ,  \hspace{2em}
g_i \to g_i\, \sqrt{(z_i^* z_i)} \ .
\ee

 Several smallest eigenvalues might be determined by applying the above
 conjugate gradient iteration subsequently to the projection into the
 orthogonal subspaces defined by
\be \label{apB_9}
H_k = P^\perp_k H P^\perp_k \ , \hspace{2em}
P^\perp_k v \equiv v - \sum_{i=1}^{k-1} v_i\, (v_i^* v) \ , \hspace{2em}
(k=2,3,\ldots) \ .
\ee
 Here $v_i$ denote the previously found normalized eigenvectors.
 This naive procedure becomes numerically instable after a few
 eigenvalues because of the numerical errors in the projectors
 $P^\perp_k$.
 One can stabilize and speed up this {\em sequential search} if one
 embeds it in an iterative scheme \cite{KALKSIMM}.
 If one is interested in the $k_{max}$ smallest eigenvalues then, after
 finding some approximation to $v_1,v_2,\ldots,v_{k_{max}}$ in a
 sequential search, the $k_{max} \otimes k_{max}$ matrix
\be \label{apB_10}
M_{ij} \equiv (v_i^* H v_j)
\ee
 is diagonalized.
 For reasonable values of $k_{max}$ this is a small problem and the
 resulting new eigenvalues and the corresponding eigenvectors
\be \label{apB_11}
v^\prime_i = \sum_{j=1}^{k_{max}} \xi^{(i)}_j v_j
\ee
 are better than $v_i$.
 Here $\xi^{(i)}$ denotes the eigenvectors of the matrix $M$.
 After this {\em intermediate diagonalization} the sequential search
 with conjugate gradient iterations is continued.

 After the restarting of the sequential search it takes some time until
 the search directions of the conjugate gradient iterations become again
 optimal.
 Therefore it is not good to insert an intermediate diagonalization too
 often, especially at later stages when the final precision is
 approached.
 In our project a good performance could be achieved if between the
 $i$-th and $(i+1)$-th intermediate diagonalization the sequential
 search was performed with $(5+10i)$ conjugate gradient iterations.
 The application of the projectors $P^\perp_k$ becomes, even for
 moderate values of $k$, quite expensive.
 Since $P^\perp_k$ projects out approximate eigenspaces of $H$, it is
 not necessary to apply it at every conjugate gradient iteration.
 Tests show that it suffices to perform the projection only in the
 intermediate diagonalizations and, say, after every 25-th conjugate
 gradient iterations.

 The optimization of the Kalkreuter-Simma algorithm pays off very well
 \cite{SPANDEREN}.
 It turns out that the number of necessary conjugate gradient iterations
 per eigenvalue is getting smaller and smaller with increasing values
 of $k_{max}$.
 Another feature is that the last few of the $k_{max}$ eigenvalues are
 slowly converging.
 As a consequence, for computing more than $k_{max}=16$ eigenvalues,
 it is advantageous to run the algorithm with $k_{max}^\prime$, say,
 5\% larger than $k_{max}$ and stop the iteration if the smallest
 $k_{max}$ eigenvalues satisfy the stopping criterion.

%%%%%%%%%%%%%%%%%%%%%%%%%%%%%%%%%%%%%%%%%%%%%%%%%%%%%%%%%%%%%%%%%%%%%%%%
\section{High performance C++ for LGT simulations}\label{appenC}

 When starting to develop the software for the DESY-M\"unster
 Collaboration we decided to take care for the reusability and
 flexibility of the code\footnote{For more informations see
 http://pauli.uni-muenster.de/\~\ spander/susy/phd.c++.ps}.
 It is well known that object oriented design and programming (OOD, OOP)
 helps to fulfill these needs \cite{GB}.
 A widely spread prejudice against OOP is bad performance.
 But new techniques like {\it expression templates} \cite{T95A,R96,VJ96}
 or {\it temporary base class idiom} \cite{SX} encouraged us to use
 an object oriented approach for the software development.
 There were serveral reasons, which led to the decision of using C++ in
 our project.
 It is the only object oriented (OO) language available for high
 performance computers and it is a high efficient OO language.
 Message Passing (MPI) and multithreading libraries (POSIX threads) are
 also usable with C++.
 With the help of templates C++ also supports generic programming
 \cite{STR97}.
 This feature allows one, for instance, to write a code template for the
 whole lattice gauge theory (LGT) simulation without specifying the
 gauge group.
 By simply providing a gauge group class which describes the basic
 functionality of the desired gauge group (SU(2), SU(3) etc), the
 compiler is able to generate code.
 Generic programming made possible to port our Super-Yang-Mills
 simulation program from SU(2) to SU(3) in less than one week.
 The only thing we had to add was a high efficient SU(3) class which
 consisted of approximately 800 lines (less than $2\%$ of the project).

 By using these techniques the efficiency of the simulation code stands
 and falls with an efficient vector class.
 A problem that arises almost always while overloading numerical
 operators of a vector class is the generation of temporary objects.
 This problem is not only limited to C++ but also to Fortran90.
 As a simple example let us consider
\be \label{apC_1}
\vec{a} =  \underbrace{
\underbrace{\vec{x}+\vec{y}}_{\vec{t_1}}+\vec{z}}_{\vec{t_2}=
              \vec{t_1}+\vec{z}} \ .
\ee
 $\vec{t_1}$ is generated in a function, which means on the stack.
 If the function is left this temporary object has to be copied away
 from the stack using the so called copy constructor.
 That means the copy constructor is used to generate a temporary copy
 from a temporary object.
 Furthermore the compiler may generate hidden temporary vectors using
 the copy constructor.

 A popular method to avoid unnecessary copying is reference counting
 \cite{COP92}.
 But due to the additional level of indirection reference counting is
 efficient only for large vectors and suited for typical vector length
 of dynamical fermion simulations.
 The basic ideas of two other solutions which are working fine for both,
 small and large vectors are
\begin{itemize}
\item {Temporary Base Class Idiom}
\vspace{-1.2ex}
  \begin{itemize}
  \item introduces an own  class TmpVector for temporary vectors,
  \item TmpVector construct/destructed shallowly,
  \item operator+(Vector \&) returns a TmpVector,
  \item operator+(TmpVector) is implemented as TmpVector+=Vector,
  \item disadvantage: four times more operators have to be overloaded.
\end{itemize}
\vspace{-1.2ex}
\item {Expression Templates}
  \begin{itemize}
  \vspace{-1.2ex}
  \item avoids temporary objects in the first place by automatically
   transforming  \\  \verb?vector u,v,w; u=v+w;?
   at compile time (more or less) into
  \begin{verbatim}
   for (int i(0); i < u.length(); ++i)
   u[i]=v[i] + w[i];
  \end{verbatim}
  \vspace{-1.2em}
  using  template meta programming (or compile time programs).
  \end{itemize}
\end{itemize}

 On one hand it is desirable to implement a class for handling gamma
 matrices.
 On the other hand it is obvious that the gamma matrix multiplication
 has to be done at compile time rather than at run time.
 Otherwise a fermion matrix multiplication would proceed at a snail`s
 pace.
 Two techniques exist to achieve this.
\begin{itemize}
\item {Lazy Evaluation  for $\mathbf{\gamma_\mu \psi}$}
  \begin{itemize}
  \item delays computation until the result is needed.
  \item processes expressions like $\chi + \gamma_\mu \psi$ in a single
   task.
  \end{itemize}
\item {Expression $\mathbf{\gamma_\mu \gamma_\nu}$}
  \begin{itemize}
  \item forces the compiler to perform this multiplication a compile
   time using template meta programming \cite{V95B}
  \end{itemize}
\end{itemize}

 To test the efficiency of different vector classes we used a
 Monte-Carlo simulation of the two dimensional $\sigma$-model.
 This is a worst case test for a vector class because the vector length
 is three.
 The administration overhead caused by the introduction of a class can
 be huge compared to the performed operation.
 Generally the difference between the class libraries are small for
 larger vector sizes.
 As one can see in tabular (\ref{peritets_tab})
 Blitz++ (Expression templates) and NumArray.h (temporary base class)
 reach comparable speed to a  hand-optimized Fortran77 implementation on
 a T3E-512/600.
 MV++ uses reference counting which is not suitable for small vector
 sizes.
 The only commercial library math.h++ surprisingly is the slowest.
%%%%%%%%%%%%%%%%%%%%%%%%%%%%%%%%%%%%%%%%%%%%%%%%%%%%%%%%%%%%%%%%%%%%%%%%
\begin{table}[ht]
 \centerline{
\begin{tabular}{|c|c|c|c|c|}
\hline
Fortran77 & Blitz++ & NumArray.h & MV++ & math.h++ \\
\hline
4.81s & 4.93s & 5.21s & 40.1s & 69.1s \\
\hline
\end{tabular}}
\vspace*{-0.7em}
\begin{center}
\parbox{15cm}{\caption{\label{peritets_tab}\em
 Runtimes of various vector classes for the simulation of the 2d O(3)
 symmetric non-linear $\sigma$-model on the T3E-512/600 with the Cray
 C++ compiler.
}}
\end{center}
\end{table}
%%%%%%%%%%%%%%%%%%%%%%%%%%%%%%%%%%%%%%%%%%%%%%%%%%%%%%%%%%%%%%%%%%%%%%%%

\begin{figure}[htb]
\centerline{\epsfig{file=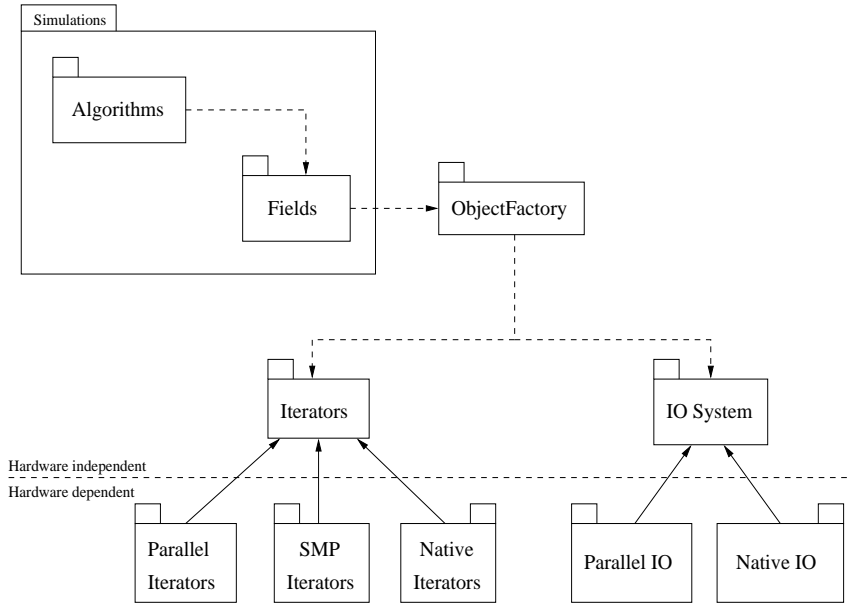,
                    width=12.0cm,height=8.0cm}}
\vspace*{0.5em}
\begin{center}
\parbox{15cm}{\caption{\label{pack_fig}\em
 UML packet structure diagram, showing the hardware dependent and
 independent parts of the project.
}}
\end{center}
\end{figure}

 Usually larger object oriented programs break up into packages which
 are only loosely connected.
 Unfortunately C++ does not support the decomposition in modules as for
 example JAVA does.
 The package structure of the simulation is shown in diagram
 \ref{pack_fig}.
 The main ingredients are algorithms which act on fields.
 They make up $90\%$ of the code.
 With the iterator pattern \cite{GHJV94} and an abstract I/O concept the
 corresponding code is hardware independent.
 It is suited for massive parallel, symmetric multiprocessor and for
 single CPU architectures.
 The hardware dependent objects, like iterators or I/O streams, should
 not be created by objects of these packages, but by the central object
 factory.

 Depending on the hardware on which the program is running, the object
 factory generates the suitable objects.
 The only hardware dependent components are the iterators and the I/O
 system.
 The question might arise why a dedicated I/O system for SMP is
 missing.
 The answer is that it is not needed.
 Only the algorithms really use more than one thread.
 When the algorithm is completed all threads are joined to a single one
 and this one uses native I/O routines.

%%%%%%%%%%%%%%%%%%%%%%%%%%%%%%%%%%%%%%%%%%%%%%%%%%%%%%%%%%%%%%%%%%%%%%%%
\newpage
\vspace*{1cm}
\begin{center}  {\Large\bf Figures}  \end{center}
\vspace*{2cm}

%%%%%%%%%%%%%%%%%%%%%%%%%%%%%%%%%%%%%%%%%%%%%%%%%%%%%%%%%%%%%%%%%%%%%%%%
%%%%%%%%%%%%%%%%%%%%%%%%%%%%%%%%%%%%%%%%%%%%%%%%%%%%%%%%%%%%%%%%%%%%%%%%

%%%%%%%%%%%%%%%%%%%%%%%%%%%%%%%%%%%%%%%%%%%%%%%%%%%%%%%%%%%%%%%%%%%%%%%%
%%%%%%%%%%%%%%%%%%%%%%%%%%%%%%%%%%%%%%%%%%%%%%%%%%%%%%%%%%%%%%%%%%%%%%%%

%%%%%%%%%%%%%%%%%%%%%%%%%%%%%%%%%%%%%%%%%%%%%%%%%%%%%%%%%%%%%%%%%%%%%%%%
%%%%%%%%%%%%%%%%%%%%%%%%%%%%%%%%%%%%%%%%%%%%%%%%%%%%%%%%%%%%%%%%%%%%%%%%

%%%%%%%%%%%%%%%%%%%%%%%%%%%%%%%%%%%%%%%%%%%%%%%%%%%%%%%%%%%%%%%%%%%%%%%%
%%%%%%%%%%%%%%%%%%%%%%%%%%%%%%%%%%%%%%%%%%%%%%%%%%%%%%%%%%%%%%%%%%%%%%%%

%%%%%%%%%%%%%%%%%%%%%%%%%%%%%%%%%%%%%%%%%%%%%%%%%%%%%%%%%%%%%%%%%%%%%%%%
%%%%%%%%%%%%%%%%%%%%%%%%%%%%%%%%%%%%%%%%%%%%%%%%%%%%%%%%%%%%%%%%%%%%%%%%

%%%%%%%%%%%%%%%%%%%%%%%%%%%%%%%%%%%%%%%%%%%%%%%%%%%%%%%%%%%%%%%%%%%%%%%%
%%%%%%%%%%%%%%%%%%%%%%%%%%%%%%%%%%%%%%%%%%%%%%%%%%%%%%%%%%%%%%%%%%%%%%%%

%%%%%%%%%%%%%%%%%%%%%%%%%%%%%%%%%%%%%%%%%%%%%%%%%%%%%%%%%%%%%%%%%%%%%%%%
%%%%%%%%%%%%%%%%%%%%%%%%%%%%%%%%%%%%%%%%%%%%%%%%%%%%%%%%%%%%%%%%%%%%%%%%

%%%%%%%%%%%%%%%%%%%%%%%%%%%%%%%%%%%%%%%%%%%%%%%%%%%%%%%%%%%%%%%%%%%%%%%%
%%%%%%%%%%%%%%%%%%%%%%%%%%%%%%%%%%%%%%%%%%%%%%%%%%%%%%%%%%%%%%%%%%%%%%%%

%%%%%%%%%%%%%%%%%%%%%%%%%%%%%%%%%%%%%%%%%%%%%%%%%%%%%%%%%%%%%%%%%%%%%%%%
%%%%%%%%%%%%%%%%%%%%%%%%%%%%%%%%%%%%%%%%%%%%%%%%%%%%%%%%%%%%%%%%%%%%%%%%

%%%%%%%%%%%%%%%%%%%%%%%%%%%%%%%%%%%%%%%%%%%%%%%%%%%%%%%%%%%%%%%%%%%%%%%%
%%%%%%%%%%%%%%%%%%%%%%%%%%%%%%%%%%%%%%%%%%%%%%%%%%%%%%%%%%%%%%%%%%%%%%%%

%%%%%%%%%%%%%%%%%%%%%%%%%%%%%%%%%%%%%%%%%%%%%%%%%%%%%%%%%%%%%%%%%%%%%%%%
%%%%%%%%%%%%%%%%%%%%%%%%%%%%%%%%%%%%%%%%%%%%%%%%%%%%%%%%%%%%%%%%%%%%%%%%

%%%%%%%%%%%%%%%%%%%%%%%%%%%%%%%%%%%%%%%%%%%%%%%%%%%%%%%%%%%%%%%%%%%%%%%%
%%%%%%%%%%%%%%%%%%%%%%%%%%%%%%%%%%%%%%%%%%%%%%%%%%%%%%%%%%%%%%%%%%%%%%%%

%%%%%%%%%%%%%%%%%%%%%%%%%%%%%%%%%%%%%%%%%%%%%%%%%%%%%%%%%%%%%%%%%%%%%%%%
%%%%%%%%%%%%%%%%%%%%%%%%%%%%%%%%%%%%%%%%%%%%%%%%%%%%%%%%%%%%%%%%%%%%%%%%

%%%%%%%%%%%%%%%%%%%%%%%%%%%%%%%%%%%%%%%%%%%%%%%%%%%%%%%%%%%%%%%%%%%%%%%%
%%%%%%%%%%%%%%%%%%%%%%%%%%%%%%%%%%%%%%%%%%%%%%%%%%%%%%%%%%%%%%%%%%%%%%%%

%%%%%%%%%%%%%%%%%%%%%%%%%%%%%%%%%%%%%%%%%%%%%%%%%%%%%%%%%%%%%%%%%%%%%%%%
%%%%%%%%%%%%%%%%%%%%%%%%%%%%%%%%%%%%%%%%%%%%%%%%%%%%%%%%%%%%%%%%%%%%%%%%

%%%%%%%%%%%%%%%%%%%%%%%%%%%%%%%%%%%%%%%%%%%%%%%%%%%%%%%%%%%%%%%%%%%%%%%%
%%%%%%%%%%%%%%%%%%%%%%%%%%%%%%%%%%%%%%%%%%%%%%%%%%%%%%%%%%%%%%%%%%%%%%%%

\end{document}